\documentclass[trackchanges,twocolumn]{aastex701}

\newcommand{\numax}{\nu_{\text{max}}}
\newcommand{\Dnu}{\Delta\nu}

\newcommand{\Teff}{T_{\text{eff}}}

\newcommand{\logg}{\log g}

\newcommand{\msun}{\mathrm{M}\textsubscript{\(\odot\)}}
\newcommand{\rsun}{\mathrm{R}\textsubscript{\(\odot\)}}

\shorttitle{Red Giant TESS Asteroseismology in NGC 188 \& NGC 6791}

\begin{document}

\title{TESS Asteroseismology of Red Giants in the Old Metal-Rich Open Clusters NGC 188 \& NGC 6791}

\author[orcid=0000-0003-0929-6541]{Madeline Howell}
\affiliation{Department of Astronomy, The Ohio State University, 140 West 18th Avenue, Columbus, OH 43210, USA}
\affiliation{Center for Cosmology and Astroparticle Physics (CCAPP), The Ohio State University, 191 West Woodruff Avenue, Columbus, OH 43210, USA}
\email[show]{howell.753@osu.edu}  

\author[orcid=0000-0001-7258-1834]{Jennifer A. Johnson}
\affiliation{Department of Astronomy, The Ohio State University, 140 West 18th Avenue, Columbus, OH 43210, USA}
\affiliation{Center for Cosmology and Astroparticle Physics (CCAPP), The Ohio State University, 191 West Woodruff Avenue, Columbus, OH 43210, USA}
\email{johnson.3064@osu.edu}  

\author[orcid=0000-0002-7549-7766]{Marc H. Pinsonneault} 
\affiliation{Department of Astronomy, The Ohio State University, 140 West 18th Avenue, Columbus, OH 43210, USA}
\affiliation{Center for Cosmology and Astroparticle Physics (CCAPP), The Ohio State University, 191 West Woodruff Avenue, Columbus, OH 43210, USA}
\email{pinsonneault.1@osu.edu}

\author[orcid=0000-0002-1333-8866]{Leslie M. Morales}
\affiliation{Department of Astronomy, University of Florida, Gainesville, FL 32611, USA}
\email{l.morales@ufl.edu} 

\author[orcid=0000-0002-4818-7885]{Jamie Tayar}
\affiliation{Department of Astronomy, University of Florida, Gainesville, FL 32611, USA}
\email{jtayar@ufl.edu} 

\author[orcid=0000-0002-2854-5796]{John D. Roberts}
\affiliation{Department of Astronomy, The Ohio State University, 140 West 18th Avenue, Columbus, OH 43210, USA}
\affiliation{Center for Cosmology and Astroparticle Physics (CCAPP), The Ohio State University, 191 West Woodruff Avenue, Columbus, OH 43210, USA}
\email{roberts.2158@buckeyemail.osu.edu}  

\author[orcid=0000-0002-4879-3519]{Dennis Stello}
\affiliation{School of Physics, University of New South Wales, NSW 2052, Australia}
\affiliation{Sydney Institute for Astronomy (SIfA), School of Physics, University of Sydney, NSW 2006 Australia}
\email{d.stello@unsw.edu.au} 

\author[orcid=0000-0002-1715-1257]{Madeleine McKenzie}
\affiliation{Observatories of the Carnegie Institution for Science, 813 Santa Barbara St., Pasadena CA 91101, USA}
\email{mmckenzie@carnegiescience.edu} 





\begin{abstract}

Open clusters are fundamental laboratories for investigating stellar and Galactic evolution, and serve as important benchmarks for asteroseismic analyses. Using a boutique method to analyze \textit{TESS} photometry, we study red giants in two old metal-rich open clusters: NGC 188 \& NGC 6791. By comparing \textit{Kepler} and \textit{TESS} observations for NGC 6791, similar oscillation mode frequencies are recovered, however we find a systematic offset of 2.2\% with a scatter of 9\% in the $\numax$ measurements. We attribute this discrepancy to the lower signal-to-noise of the \textit{TESS} data for these relatively faint stars. For the brighter cluster NGC 188, we present new seismic measurements in 17 red giants. We estimate average seismic masses for the RGB of $M_{\text{RGB,NGC188}} = 1.13\pm0.04$(rand)$^{+0.12}_{-0.19}$(sys)$~\msun$ and RC of $M_{\text{RC,NGC188}} = 1.11\pm0.01$(rand)$^{+0.11}_{-0.19}$(sys)$~\msun$, consistent with independent mass estimates for this cluster and with similar precision to previous \textit{Kepler} studies. From the difference between the average evolutionary phase masses, we estimate an integrated RGB mass loss of $\Delta M = 0.02 \pm 0.04$(rand)$\pm0.01$(sys)$\msun$, supporting the evidence for lower mass loss at higher metallicities. Using asteroseismology and chemical abundances, we identify three binary interaction candidates: two under-massive stars and one over-massive star potentially exhibiting dipole-mode suppression. Finally, we derive an average seismic cluster age of $7.0\pm0.9$~Gyrs, in good agreement with previous literature ages. Our analysis demonstrates the strong potential of \textit{TESS} asteroseismology for open clusters, and motivates extending this investigation to other \textit{TESS} clusters that span a wider range of ages and metallicities.

\end{abstract}



\section{Introduction} 
\label{sec:intro}

Open clusters are essential testbeds for stellar and Galactic evolution, because their stars share a common distance and are homogeneous in metallicity and age. By deriving accurate stellar masses and ages, clusters can be used to provide critical constraints on physics used in stellar modeling \cite[e.g.][]{Reyes25_Nature} and to trace the chemical evolution of the Milky Way discs, known as Galactic archaeology \citep[e.g.][]{Myers2022_OCCAMDR17,Otto25_OCCAMDR19}. Through the detection of stochastic pressure-mode oscillations that propagate in the convective envelopes of stars, asteroseismology is emerging as a gold standard method to precisely measure stellar masses, radii, and ages for red giants using time-series photometry from space-based missions. Open clusters provide a benchmark for these asteroseismic analyses, as we can determine the accuracy of the seismic masses by directly comparing to independent mass scales, such as isochrones fits to cluster photometry. Furthermore, we can study a population of stars that share homogeneous cluster properties in differing evolutionary phases to test stellar evolution theories.

Individual stellar masses and radii of red giants can be quantified using asteroseismic scaling relations. This involves measuring two `global' seismic quantities: the frequency of the maximum acoustic power, $\nu_{\text{max}}$, which is related to the surface gravity and effective temperature, and the large frequency spacing, $\Delta\nu$, which is related to the mean density of the star \citep{Ulrich1986_scaling_relation1,Brown1991_scaling_relation2,Kjeldsen1995_scaling_relation3}. Individual masses derived from the seismic scaling relations can then be used to infer seismic ages from isochrones. This has been accomplished for large samples of field red giants using \textit{Kepler} and \textit{K2} time-series photometry \citep{Schon-Stasik24_APOK2,Pinsonneault2025_APOKASC3}. The combination of stellar properties derived from asteroseismology with precise Gaia astrometry, and spectroscopic chemical abundances and stellar parameters, have culminated in large surveys of well-characterized red giant stars. The seismic radii are typically calibrated to the Gaia radius scale \citep{Zinn19_radiuscalibrations}, however it is more difficult to assess the accuracy of the seismic masses. This is mitigated for seismic measurements of open cluster red giants, as their distances are well-constrained and we can directly compare to isochrones. 

Galactic archaeology is being revolutionized by the ability to determine asteroseismic ages for large samples of red giant stars \citep[e.g.][]{Pinsonneault2025_APOKASC3,Theodoridis2025}. However, ages inferred from asteroseismology of red giants have only been derived for a few open clusters using \textit{Kepler/K2} photometry; this includes NGC 6866 \citep{Brogaard23_NGC6866}, M67 \citep{Reyes25_M67}, NGC 1817, NGC 6819 \& NGC 6791 \citep{Pinsonneault2025_APOKASC3,Tayar25}. This limited sample is largely a consequence of the restricted fields of view of the \textit{Kepler} and \textit{K2} missions, and the small number of open clusters with substantial red giant populations. The all-sky \textit{TESS} mission \citep{Ricker15_TESS} provides an unprecedented opportunity to seismically characterize red giants in a larger number of open clusters, and measure the seismic ages for a greater sample of these objects. However, \textit{TESS} photometry has a higher likelihood of photometric contamination from neighboring stars in the densely populated regions of clusters due to the instrument's larger pixels compared to \textit{Kepler}. Additionally, there can be systematics in the \textit{TESS} light curves caused by scattered Earth and moon light that are not always removed by automated photometric pipelines. Therefore, careful analysis of the \textit{TESS} data is needed to isolate clean photometric light curves for single stars. 

Our study uses a boutique photometric analysis to measure the seismic quantities in two old and metal-rich open clusters: NGC 188 \& NGC 6791. Both clusters have been observed for several months by the \textit{TESS} mission, allowing for asteroseismic analyses of their red giant populations across two evolutionary phases: red giant branch (RGB) and core-helium burning red clump (RC) stars. NGC 6791 was also observed in the \textit{Kepler} nominal mission, and therefore there already exists precise asteroseismic masses and ages for cluster members in the literature \citep{Basu2011_NGC6791,Miglio2012_OCs,Brogaard6791_undermassivestar,CoveloPaz23_NGC6791,Pinsonneault2025_APOKASC3,Ash25}. Rather than re-deriving these quantities, we instead aim to directly compare the quality of the \textit{Kepler} vs \textit{TESS} photometry for members of this cluster. NGC 188 has no existing seismic analyses in the literature, and hence this work will present new asteroseismic measurements. From the seismic masses, we aim to confirm the cluster's old age, and measure an integrated mass loss to contribute to further analyses on the mass loss-metallicity trend \citep[e.g.][]{Li25_massloss}. Additionally, by coupling the seismically-derived parameters with chemical abundances, we can identify stars that have experienced a non-standard evolutionary path \citep[e.g.][]{Bufanda23,Frazer25,Roberts26}. 

Field populations have a range of ages, making it difficult to connect RGB stars to their RC descendants directly. Therefore, clusters offer a crucial advantage over field stars for studying stellar mass loss; the mass lost during the RGB phase can be directly inferred by comparing stellar masses on the RGB with those in the subsequent RC phase. Measuring this evolutionary process is important since the physical mechanism underlying RGB mass loss in low mass stars remains poorly understood. Empirical mass loss prescriptions, such as Reimers' \citep{Reimers1975_massloss_rate} or Schr\"oder-Cuntz \citep{SchroderCuntz05}, are simple relations dependent on stellar properties and are commonly used to estimate the mass loss rates in stellar models. These relations predict a weak correlation between the total mass lost on the RGB and metallicity \citep[e.g.][]{Li25_massloss,Roberts26}. This behavior is consistent with the asteroseismic integrated mass loss estimates for four metal-poor globular clusters \citep[][]{Howell25_pyMON}, and model fits to color-magnitude diagrams \citep[CMDs; e.g.][]{Tailo20_massloss_difference_multipops}. Measurements of an integrated mass loss have also been quantified for the thick disk, where there is a well-defined mean mass-age-metallicity relationship, and the thin disk, where a field turnoff sets minimum masses for the RGB and RC populations \citep[e.g.][]{Miglio21,Li25_massloss}. Combining these field star estimates with asteroseismic integrated mass loss measurements from open clusters have shown an opposing trend to the metal-poor regime; metal-rich stars have a smaller measured integrated mass loss compared to metal-intermediate stars for the metallicity range [Fe/H]~$\gtrsim -0.8$ \citep{Brogaard24_massloss,Li25_massloss,Marasco25_massloss}. The conflicting relations between the old metal-poor stars and younger metal-rich stars could be interpreted in terms of the differences in initial mass and shorter lifetime on the RGB. However, this explanation cannot describe the behavior in the metallicity range $-0.8\lesssim$~[M/H]~$\lesssim -0.5$, where the two trends overlap but do not fully agree \citep{Li25_massloss}. The apparent dichotomy in the mass loss-metallicity trends remains an open question, where additional measurements of the integrated mass loss is needed. It is difficult to test mass loss in intermediate-aged field populations, whereas open clusters are ideal laboratories for this purpose. Deriving the RGB integrated mass loss in the Solar-metallicity cluster NGC 188 tests the behavior of the metal-rich mass loss relationship, and could motivate further studies of clusters observed by \textit{TESS} that are close to or within the contentious metallicity range. 

Through asteroseismic surveys of red giants in stellar clusters, we can also identify stars with atypical masses thought to be the consequence of non-standard evolutionary pathways. Binary interaction events leading to over-massive stars are common for intermediate to old open clusters, which is evidenced by the distinct population of stars that are more luminous and hotter than the main sequence turn off in a cluster CMD \citep{Sandage1952_BSSCMD}. These objects are referred to as blue straggler stars while on the main sequence, and are thought to form from either merger \citep[e.g.][]{Mateo1990_BSS,Chen2008_BSS,Wang2022_BSS} or mass transfer \citep{McCrea1964_BSS} events. On the contrary, a population of under-massive sub-subgiants have also been detected in open clusters \citep[][]{Belloni1998}, which could be the remnants of mass stripping events \citep[e.g.][]{Mathieu2003,Leiner2017}. After exhausting their core hydrogen, both the blue straggler stars and sub-subgiants evolve off the main sequence and ascend the RGB, where their evolutionary tracks converge towards a typical cluster member's isochrone \citep[e.g.][]{Handberg2017,Leiner2017}. As such, they can become photometrically indistinguishable from the cluster's main red giant population. For these phases of evolution, we need an alternative diagnostic for identifying merger or mass transfer candidates. Asteroseismology allows us to search for RGB and RC stars with outlying masses, which is a powerful indicator of an atypical evolutionary event. This has been previously achieved in open and globular clusters, with the discovery of both under-massive and over-massive stars that deviate significantly from the average mass for their respective evolutionary phases \cite[e.g.][]{Handberg2017,Brogaard6791_undermassivestar, Howell22_M4,Reyes25_M67}. We aim to extend these studies by searching for anomalous asteroseismic masses in the NGC 188 red giant population.



This paper is structured as follows: Section~\ref{sec:target_selection} details the selection of cluster member candidates for asteroseismic analyses, Section~\ref{sec:photometric_analysis} presents the method of analyzing the \textit{TESS} photometry, Sections~\ref{sec:seismic_analysis} \& \ref{sec:stellar_parameters} report the method of measuring the seismic quantities and stellar parameters that are used to estimate the masses and ages in Sections~\ref{sec:mass_results} \& \ref{sec:age_results}. Finally, we expand on the asteroseismic analysis of a sub-sample of NGC 188 stars in Section~\ref{sec:gold_sample}, and provide our conclusions in Section~\ref{sec:conclusion}.

\section{Target Selection}
\label{sec:target_selection}

\subsection{Magnitude Limit for Detecting Oscillations with TESS photometry}
\label{sec:mag_limit}
Previous asteroseismic studies of large red giant samples in open clusters predominantly used photometry from the \textit{Kepler} missions. The combination of \textit{Kepler}’s 4" pixel scale and the observing baseline of several years for the nominal mission and 80-days in the second mission allowed for precise measurements of seismic masses and ages across a large range in magnitude and for a large sample of stars. For example, the APOKASC-3 catalog \citep{Pinsonneault2025_APOKASC3} reported detections of solar-like oscillations in 53 red giant stars in NGC 6791, spanning from approximately 1.5 magnitudes fainter than the RC to the RGB tip. Directly replicating these results with the \textit{TESS} mission is unlikely due to the larger pixel sizes (21"/pixel), and shorter observing windows split into 27-day sectors. Consequently, this reduces the magnitude range where asteroseismic oscillations can be detected using \textit{TESS}. Before compiling candidate samples, we first estimated the detection magnitude limits for oscillations in both of our target clusters.

\begin{figure*}[ht!]
\plotone{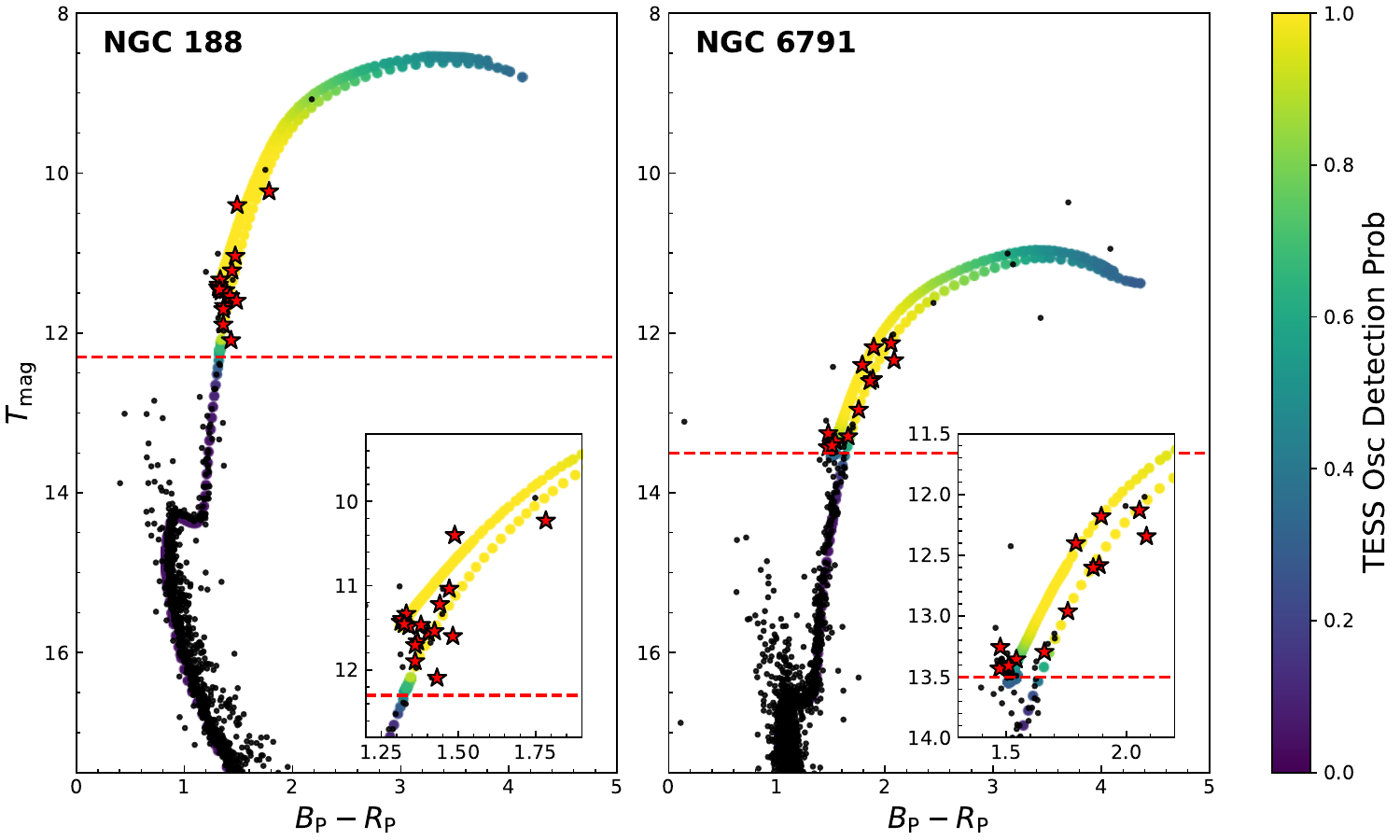}
\caption{CMDs for NGC 188 (left) and NGC 6791 (right) using the \textit{TESS} magnitude, $T_{\rm mag}$, and Gaia DR3 colour \citep{GaiaDR3_releasepaper}, ($B_{P} - R_{P}$). A PARSEC isochrone (see Table~\ref{tab:isochrones} for input parameters) is colored by the oscillation detection probabilities. The red dashed line denotes where the oscillation detection probability is 50\%. The cluster membership sample (black) is from \citet{Hunt2024_ocmembership} and our seismic sample is shown by red star symbols. The insets present a zoomed-in region of the colour–magnitude diagram containing our seismic sample.
\label{fig:cluster_CMDs}}
\end{figure*}

\citet{Chaplin2011_oscdetectionprobs} developed a method to calculate the probability of detecting asteroseismic oscillations with the \textit{Kepler} instrument. This approach estimates a signal-to-noise (SNR) ratio from the predicted power of the oscillations, given an observation length, magnitude, and $\numax$, and a noise model that incorporates components from both the instrumental and granulation noise. This method was later adapted for the \textit{TESS} instrument by \citet{Campante2016_tessoscdetectionprobs} and \citet{Hey2024_ATLcode}, and has been used to construct asteroseismic target lists for the \textit{TESS} mission \citep{Schofield2019_ATL}. We used the \texttt{tess-atl} python package \citep{Hey2024_ATLcode} to calculate the oscillation detection probabilities for each cluster. The calculation of this metric requires the following inputs: \textit{TESS} magnitude ($T_{\rm mag}$), the effective temperature ($\Teff$), radius ($R$), surface gravity ($\logg $), and the number of \textit{TESS} sectors ($N_{\text{TESS}}$). To estimate these stellar parameters, we fit \texttt{PARSEC} v2.0 isochrones \citep{Bressan12_PARSEC,Nguyen22_PARSECv2, Nguyen25} to each cluster (see Table~\ref{tab:isochrones} for isochrone input parameters). The number of observed \textit{TESS} sectors per cluster was determined at the position of the cluster centre using the \texttt{TESS-Point} tool \citep{Burke20_tesspoint}, resulting in 13 sectors for NGC 188 and 9 for NGC 6791. Figure~\ref{fig:cluster_CMDs} demonstrates the oscillation detection probabilities across the cluster CMDs. We adopt a minimum detection probability of 50\%, which corresponds to a $T_{\rm mag}$ limit of $12.3$ for NGC 188 and $13.5$ for NGC 6791.


\begin{deluxetable*}{lcc}
\digitalasset
\tablewidth{0pt}
\tablecaption{In the first segment of the table, we provide the input parameters to PARSEC isochrones. The second segment reports the predicted initial mass at the RGB bump ($M_{\text{RGB bump}}$) and zero-age RC ($M_{\text{RC}}$), and the final segment reports the current mass at the same evolutionary points from the isochrones. \label{tab:isochrones}}
\tablehead{
\colhead{Input Parameters} & \colhead{NGC 188} & \colhead{NGC 6791}   }
\startdata
[M/H] & 0.08 & 0.3 \\
Age (Gyrs) & 6.5 & 8.3 \\
Distance (pc) & 1790 & 3750 \\
E(B-V) (mag) & 0.07 & 0.18 \\
Reimers' $\eta_{\text{R}}$ & 0.1 & 0.1 \\
\hline
$M_{\text{RGB bump, initial}}$ ($\msun$) & 1.18 & 1.14 \\
$M_{\text{RC,initial}}$ ($\msun$) & 1.19 & 1.15 \\
\hline
$M_{\text{RGB bump,current}}$ ($\msun$) & 1.18 & 1.13 \\
$M_{\text{RC,current}}$ ($\msun$) & 1.11 & 1.06 \\
\enddata
\end{deluxetable*}

\subsection{Cluster Membership}
\label{sec:membership}
Target samples were originally selected from the \citet{Hunt2024_ocmembership} open cluster membership catalog, which uses \textit{Gaia} DR3 astrometry and hierarchal modelling to calculate membership probabilities. Stars fainter than the limiting magnitude for asteroseismic detection probabilities (see Section~\ref{sec:mag_limit}) were excluded from these samples, resulting in initial samples sizes of 26 stars for NGC 188 and 35 stars in NGC 6791. We confirm the membership classification by checking for homogeneity in the spectroscopic radial velocities and metallicities from APOGEE DR17 \citep{Abdurrouf22_APOGEEDR17}, if available. Following the criterion used by the sixth release of the Open Cluster Chemical Abundances and Mapping (OCCAM) Survey \citep{Myers2022_OCCAMDR17}, a star is considered a member if their radial velocity and metallicity lies within $3\sigma$ of the cluster mean.


There were five stars in NGC 188 and six stars in NGC 6791 that did not have spectroscopic measurements in the APOGEE DR17 survey. For this sub-sample, we performed our own spectroscopic membership check using the metallicity measurements inferred from Gaia XP spectra using the XGBoost algorithm \citep{Andrae2023_GaiaXGBoost}. Since this catalog was trained on the APOGEE DR17 catalog, the metallicities are consistent with the remainder of our sample. We note that the radial velocities provided in the Gaia XP catalog are from the Gaia DR3 catalog and were already incorporated in the \citet{Hunt2024_ocmembership} membership study. Therefore, we only consider the metallicity for the spectroscopic membership classification of these stars. We use the cluster average metallicity and uncertainty from the OCCAM survey ([Fe/H]$_{\text{NGC188}}=0.07\pm0.04$~dex \& [Fe/H]$_{\text{NGC6791}}=0.31\pm0.04$~dex) to distinguish members from the Gaia XP samples. We found three stars in NGC 188 and one star in NGC 6791 that did not meet the metallicity membership criteria, and were consequently removed from our study. Our final member samples consisted of 23 stars in NGC 188 and 34 stars in NGC 6791, which will comprise the candidates for asteroseismic analysis.

\subsection{Evolutionary Stage Classification}
\label{sec:evol_classification}

We classified our member samples into the RGB and RC using Eqs.~10~\&~11 from \citet{Holtzman18}. This spectroscopic method involved assigning an evolutionary label based on the star's relative position to a ridge-line that was fitted to the RGB in $\Teff$-$\logg$ space, and the metallicity and [C/N] abundance. If the star did not have a carbon or nitrogen measurement, or the [C/N] abundance was inconsistent with the cluster average\footnote{It has been shown that the [C/N] measurement for a mono-age population is the same on the RGB after first dredge up and in the RC for metal-rich stars. See \citet{Roberts25} for more details.}, then it was classified from Gaia DR3 photometry only. For the NGC 6791 sample, we adopt the seismic evolutionary phase labels from the APOKASC-3 catalog where available. 

\section{Analysis of the TESS Photometry}
\label{sec:photometric_analysis}
The crowded stellar environments of clusters, combined with the large pixels of the \textit{TESS} instrument, makes it challenging to extract light curves for single stars. To minimize contamination from nearby sources, we extracted photometry directly from the \textit{TESS} full-frame images (FFIs), rather than relying on pipeline-generated light curves with pre-defined aperture masks. The FFIs for each star was obtained using the \texttt{TESSCut} \citep{Brasseur2019_tesscut} tool, which is implemented within the \texttt{LightKurve} \citep{LightKurve2018} python package. As a first step, we inspected the FFIs centered on the target star, with the positions of nearby Gaia DR3 sources overlaid to identify potential contaminants. We define contaminants as stars: i) that are located within a radius of two pixels of the target star, and ii) have a Gaia G-magnitude that is within $\pm2$ magnitudes of the target star. For uncontaminated candidates, we performed custom aperture photometry using manually defined aperture masks. We refer to this as a `boutique' method of generating light curves, since the aperture masks were tailored individually for each target and each sector. The choice of aperture mask was guided by any potential contaminants located outside of the 2-pixel radius and visual inspection of a preliminary periodogram of the light curve to search for a solar-like oscillation signal with the highest SNR. All aperture masks were comprised of a minimum of $2\times2$ pixel grid.

We detrended the photometry extracted from the custom aperture masks for each sector individually using the following steps\footnote{Our detrending pipeline can be accessed at \href{https://github.com/maddyhowell/BELA-TESS}{https://github.com/maddyhowell/BELA-TESS}}:
\begin{itemize}
    \item Homogeneous time sampling of light curves among sectors:  Throughout the duration of the \textit{TESS} mission, the rate of sampling (known as the cadence) of the FFI images increased. In the prime mission (sectors 1-26) the FFI cadence was 30 minutes, for the first extended mission (sectors 27-55) the cadence was 10 minutes, and for the second extended mission (sectors 56-current) the cadence was 200 seconds. For sectors $>$ 26, we binned the time-series photometry to 30 minutes, using an inbuilt function in \texttt{LightKurve} to ensure there is homogeneous sampling between sectors.
    \item Correcting for background noise: We used the \texttt{LightKurve} background-correction tool \texttt{RegressionCorrector}, which was designed to remove scattered light from the Earth and Moon in the \textit{TESS} FFIs. This method reduces the flux in pixels surrounding the defined aperture mask to 7 vectors via principal component analysis. This ensures that noise due to stochastic or periodic variability in the background pixels is attenuated. A background noise model is then constructed as the optimal linear regression to the 7 principal components, and is subtracted from the target light curve.
    \item Outlier removal: We reject spurious outlying flux artifacts that deviate by more than $3\sigma$ from the median flux.
    \item Filling gaps in the time-series photometry: We filled gaps that were less than 1.5 hours using a linear interpolation. Removing short gaps has been shown to reduce high-frequency noise in red giant stars due to spectral leakage \citep{Stello2015}.
    \item High-pass filter: A high-pass triangular filter was applied with a filter width of 4 days to remove the long-term trends within the time-series photometry. For RGB stars with a spectroscopic $\logg \lesssim 2.2$ (corresponding to a predicted $\numax \lesssim 20 \mu$Hz), we used a filter of 12 days to avoid attenuating the oscillation power at lower frequencies. This filtering step normalizes the light curves in units of parts per million (ppm).
\end{itemize}

We ignored sectors of the light curves that still displayed anomalous features that had not been removed via our detrending process. This included obvious trends from background noise that were not captured in our background-corrector step or signatures from eclipsing binaries. Additionally, if the star was situated close to the edge of the camera in a sector, it was excluded. This resulted in at most three sectors being removed per star for NGC 188, where the final light curves consisted of $\ge10$ \textit{TESS} sectors for each star (see table in Appendix~\ref{sec:A1_ParamTables}). We had to remove more sectors for some of our fainter NGC 6791 sample, with the total numbers of sectors used per star ranging between 2-7 (Appendix~\ref{sec:A1_ParamTables}). This was partially due to the higher occurrence of stars being situated at the edge of a camera, which sometimes happened in up to three sectors. 

The remaining sectors were combined using the \texttt{stitch} function in \texttt{LightKurve}, producing the final full light curve for each target. We chose to keep the large gaps in between sectors to preserve the phase information of the oscillation signal and to avoid introducing artificial structures in the power spectra \citep[see][for more information]{Bedding2022}. We calculated the power spectral density using a Lomb-Scargle periodogram \citep{Lomb,Scargle}, which included an over-sample factor of 5. The frequency resolution for the Fourier transform was determined from the total observation time (i.e. excluding gaps). This ensures that the frequency resolution was not under-estimated leading to fine structures in the power spectrum that increased the overall noise. We inspected the resulting power spectra for an asteroseismic signal, using a prediction for $\numax$ calculated from a spectroscopic $\Teff$ and $\logg$ ($\numax\propto g\Teff^{-0.5}$) to guide our search. 

In total, we detect solar-like oscillations for 17 stars in NGC 188 and 12 stars in NGC 6791. The primary reason for not detecting an asteroseismic signal in the remaining candidates was contamination; these stars were removed from our target samples prior to the light curve analysis stage following our contamination criteria. Our final seismic samples are indicated by the red star symbols in Figure~\ref{fig:cluster_CMDs}, demonstrating that asteroseismic signals are detected to our limiting \textit{TESS} magnitude corresponding to a 50\% detection probability using the \citet{Chaplin2011_oscdetectionprobs} method. As a test, we searched for asteroseismic signals in a fainter sample of stars. This sub-sample includes stars with \textit{TESS} magnitudes as faint as $T_{\mathrm{mag}} = 12.5$ for NGC 188 and $T_{\mathrm{mag}} = 13.8$ for NGC 6791, corresponding to a detection probability of $\sim25\%$. Weak evidence of solar-like oscillations were seen in five NGC 6791 giants, however the SNR for each star was too low to get a reliable $\numax$ measurement. Hence, we conclude that a limiting detection probability of 50\% using the \citet{Chaplin2011_oscdetectionprobs} method was sufficient for our study.

\begin{figure}[ht!]
\plotone{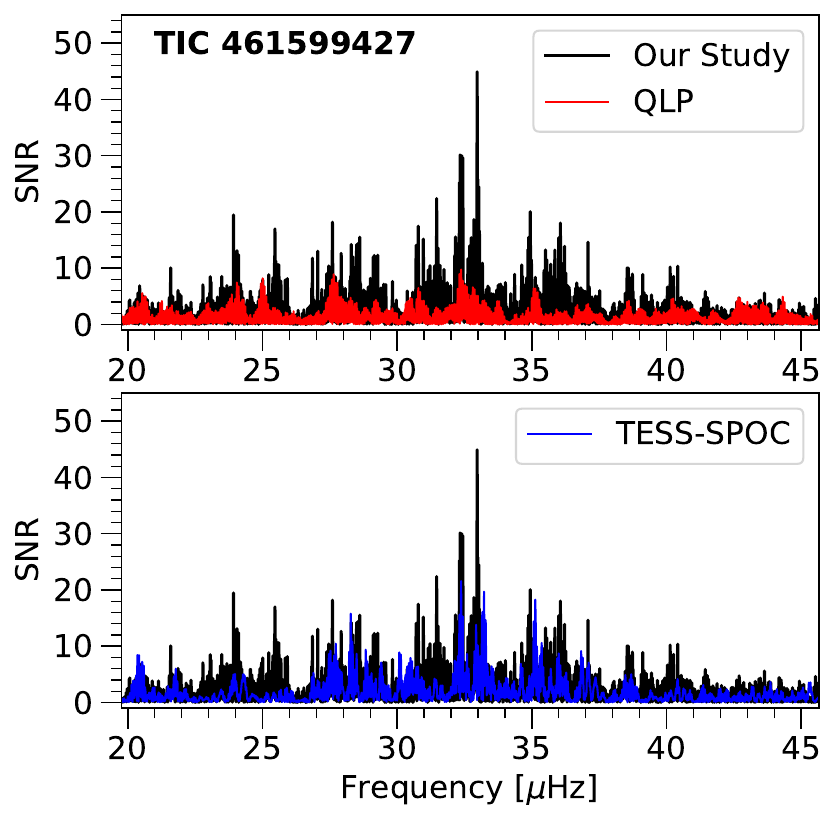}
\caption{Example background-corrected power spectra for a NGC 188 star (TIC 461599427) demonstrating the increase in SNR when using a boutique method of producing light curves (black) compared to a pipeline generated light curves. The top panel shows a comparison to QLP light curve, and the bottom panel shows a comparison to the TESS-SPOC light curve. 
\label{fig:QLPvsboutique}} 
\end{figure}

In addition to having custom aperture masks to avoid contamination, our boutique method enables a comprehensive analysis of the entire set of \textit{TESS} observations per star by starting from the FFI data. Whereas, pipeline generated light curves are not often produced for every observed sector or every star. Our careful treatment of the \textit{TESS} photometry results in a larger SNR in our power spectra when compared to pipeline generated light curves. This is demonstrated in Figure~\ref{fig:QLPvsboutique} for an example background-corrected power spectrum for a NGC 188 star (TIC 461599427) when compared to a `Quick Look Pipeline' (QLP)  light curve (top panel; \citealt{Huang2020_QLP,Huang2020B_QLP,Kunimoto2021_QLP,Kunimoto22_QLP}) and a TESS-SPOC light curve (bottom panel; \citealt{Jenkins_TESSSPOC,Caldwell_TESSSPOC}). The QLP light curve method adopts an aperture mask based on the $T_{\text{mag}}$ of the star. This automated aperture mask selection could contribute to the decrease in SNR in their power spectrum compared to our method, as light from contaminating nearby sources or background noise could be included in their photometry. The QLP has been executed for all \textit{TESS} sectors for TIC 461599427, hence has a similar observing baseline to our light curve. In contrast, the TESS-SPOC light curves are less widely available, where only three sector light curves were provided for our example star. Despite the smaller observing baseline, the TESS-SPOC power spectrum has a larger SNR in comparison to the QLP power spectrum, but smaller compared to our boutique power spectrum. We acknowledge that our boutique method is not ideal for large asteroseismic samples, however our analysis demonstrates that careful individualized treatment of all the available photometry for smaller samples of \textit{TESS} targets is recommended, rather than adopting pipeline generated products. 


\section{Asteroseismic Analysis}
\label{sec:seismic_analysis}

Following other asteroseismic red giant studies using \textit{TESS} photometry \citep[e.g.][]{Hon21_tessrgbcatalog}, we initially only measure the frequency of maximum acoustic power, $\numax$. Using a $\numax$-radius seismic scaling relation has been shown to yield highly accurate asteroseismic masses and ages when benchmarked against cluster-calibrated stellar models \cite[e.g.][]{Howell22_M4, Howell24_M80, Reyes25_M67}. To measure $\numax$, we used the \texttt{pyMON} pipeline \citep{Howell25_pyMON}, which is directly based on the asteroseismic pipeline \texttt{pySYD} \citep{Chontos21_pySYD}. The \texttt{pyMON} pipeline is optimized for stars where it can be difficult to measure an accurate large frequency spacing ($\Delta\nu$); for example stars that oscillate at low-$\numax$ values where there is a reduction in the number of radial orders \citep{Stello2014_lownumaxstars}, and for lower SNR stars. Instead of directly measuring the $\Delta\nu$ quantity from the spectrum, \texttt{pyMON} assumes a large frequency spacing from a $\Delta\nu$-$\numax$ scaling relation \citep[in the form $\Delta\nu\propto\alpha\numax^{\beta}$, where $\alpha$ and $\beta$ are constants; see][]{Stello09_dnu_numax_relation}. The $\Delta\nu$ is used to smooth a granulation noise-corrected power spectrum, where $\numax$ is measured as the frequency corresponding to the largest amplitude within the oscillation envelope. Example power spectra are demonstrated in Figures~\ref{fig:QLPvsboutique}~\&~\ref{fig:NGC6791_snrplots}, and Appendix~\ref{sec:A3_ModeVisibilties}. The $\numax$ measurements are reported in Appendix~\ref{sec:A1_ParamTables}.

During our asteroseismic analyses, we identified a subset of stars in NGC 188 with particularly clear oscillation patterns. These stars are classified as our ‘gold’ asteroseismic sample, and we extended their asteroseismic analysis further in Section~\ref{sec:gold_sample}.

\subsection{Comparison of the $\numax$ measurements from TESS \& Kepler for NGC 6791}
\begin{figure}[ht!]
\plotone{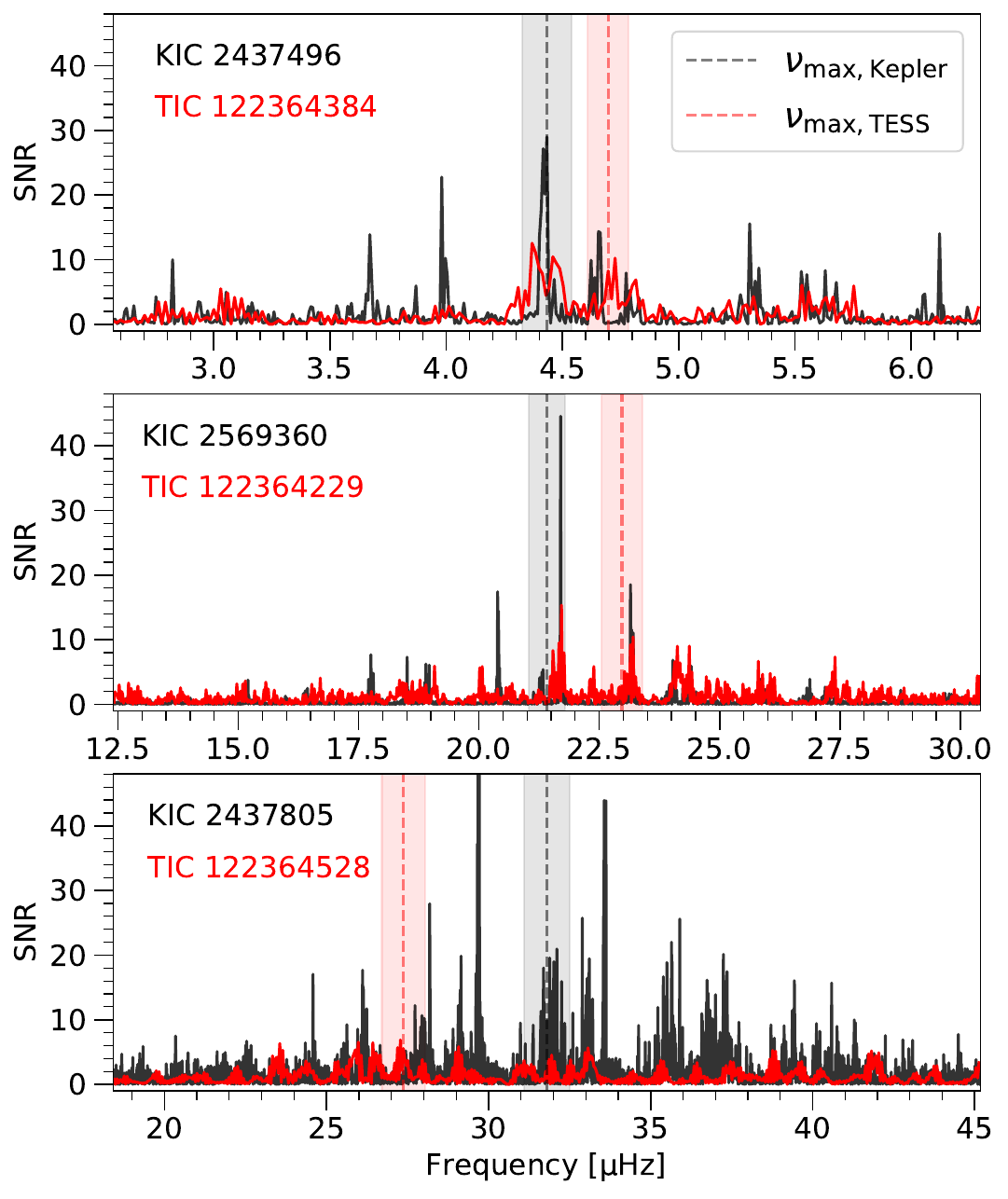}
\caption{Comparison of the \textit{Kepler} (red) and \textit{TESS} (black) background corrected power spectra (SNR) for three stars in NGC 6791. KIC and TIC IDs are annotated. The \textit{Kepler} power spectra are from the \texttt{KEPSEISMIC} database. The measured $\numax$ from the \textit{Kepler} photometry and this study are indicated by the vertical dashed lines, and the associated $1\sigma$ uncertainty by the shaded regions. The SNR of KIC 2569360 has been scaled down by a factor of 5 for visualization purposes.
\label{fig:NGC6791_snrplots}}
\end{figure}

NGC 6791 is recognized as a `\textit{Kepler} cluster', where many of the its red giant members already have high-precision asteroseismic measurements and well-constrained stellar parameters using photometry from the nominal \textit{Kepler} mission \cite[e.g.][]{Corsaro2012_OCs,Miglio2012_OCs,Brogaard6791_undermassivestar,CoveloPaz23_NGC6791,Pinsonneault2025_APOKASC3, Ash25}. In contrast, \textit{TESS} has lower spatial-resolution due to its larger pixel sizes and a significantly shorter observing baseline for stars in this cluster. Consequently, the \textit{TESS} power spectra exhibit reduced SNR compared to \textit{Kepler}, as demonstrated in Figure~\ref{fig:NGC6791_snrplots} for three representative giants observed by both missions. Despite this reduction in SNR, similar oscillation modes are still detectable from both photometric datasets, although the agreement degrades for fainter giants (i.e. with increasing $\numax$ as shown in the bottom panel). Therefore, our analysis of NGC 6791 centers on evaluating the accuracy of the $\numax$ measurements from \textit{TESS} power spectra when compared to \textit{Kepler}, rather than re-deriving stellar parameters. This work extends the \textit{Kepler}-\textit{TESS} comparison by \citet{Stello2022_KeplervTess}, which used 1-2 \textit{TESS} sectors, by testing whether longer observing baselines from additional \textit{TESS} sectors lead to improved accuracy in the \textit{TESS} measurements. 

\begin{figure}[ht!]
\plotone{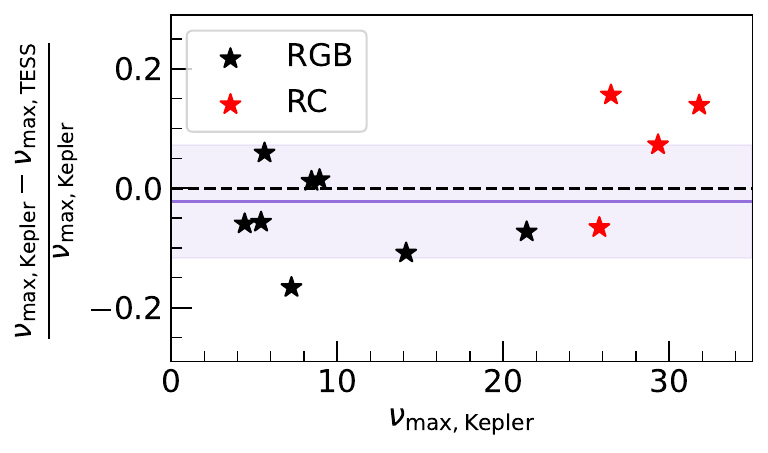}
\caption{Difference between our measured $\numax$ from \textit{TESS} photometry and the $\numax$ from \textit{Kepler} photometry for NGC 6791 oscillating RGB (black) and RC (red) stars. The  purple line and shaded region indicates the median fractional offset and scatter. A typical $\numax$ uncertainty is approximately the size of the symbol's width.
\label{fig:NGC6791_numaxcomp}}
\end{figure}

Figure~\ref{fig:NGC6791_numaxcomp} shows the differences between our $\numax$ measurements from \textit{TESS} and those derived from \textit{Kepler} for our NGC 6791 sample. The \textit{Kepler} $\numax$ values are estimated from the \texttt{KEPSEISMIC} power spectra \citep{Garcia2011_KEPSEISMIC,Garcia2014_KEPSEISMIC,Pires2015_KEPSEISMIC} using the same method as our \textit{TESS} data. This ensures that we reduce the known discrepancies between different asteroseismic pipelines (see \citealt{Pinsonneault2025_APOKASC3} for further details). We find a median fractional offset of $2.2$\% between the \textit{TESS} and \textit{Kepler} $\nu_{\text{max}}$ measurements, with a scatter of 9\%. This scatter is larger than the 5–6\% reported for \textit{Kepler} field stars in \citet{Stello2022_KeplervTess}. The difference in the scatter is likely driven by the faintness of the NGC 6791 cluster, where the sample studied in \citet{Stello2022_KeplervTess} had \textit{TESS} magnitudes of $T_{\text{mag}}\lesssim12.5$, which is brighter than majority of our NGC 6791 seismic sample. To more robustly assess the effect of extending the \textit{TESS} observing baseline on the accuracy of the $\nu_{\text{max}}$ measurements compared to \textit{Kepler}, brighter open clusters can be considered. Two examples include NGC 6819 (observed in 11 \textit{TESS} sectors) or M67 (observed in 5 \textit{TESS} sectors).

 
In contrast to NGC 6791, the cluster NGC 188 is significantly brighter (as shown in Fig.~\ref{fig:cluster_CMDs}), however was not observed by \textit{Kepler}. The remainder of this study focuses on testing whether we can use \textit{TESS} photometry to quantify reliable measurements for $\numax$ in our NGC 188 candidate sample, to enable precise estimates of the seismic masses and ages for oscillating giants. 

\section{Stellar Parameters for NGC 188}
\label{sec:stellar_parameters}

To determine masses from the seismic scaling relations, we require estimates for $\Teff$ and stellar radius for our NGC 188 sample. All oscillating giants have spectroscopic $\Teff$ measurements from either the APOGEE DR17 survey or the Gaia XP catalog. We adopt these values as the effective temperature for each star, and assume an uncertainty of $50$~K \citep{Pinsonneault2025_APOKASC3}. Using spectroscopic rather than photometric $T_{\rm eff}$ estimates (e.g. from the infrared flux method) helps mitigate systematics associated with dust extinction. However, extinction-related effects in the seismic masses cannot be entirely removed, as the Gaia-based radius calculation is dependent on a dust correction term. 

There is broad agreement in the literature that NGC 188 has a low line-of-sight dust extinction. However, the reported estimates for the average extinction in the visual band ($A_{V}$) varies widely across studies (see literature summary in \citealt{Cinar24_NGC188SED}), which can introduce systematics into the seismic masses as discussed further in Sec.~\ref{sec:systematic_uncertainties}. For example, two recent studies measured discrepant averages of $A_{V}$ of $0.11\pm0.09$~mag \citep{Cinar24_NGC188SED} and $0.34\pm0.04$~mag \citep{Yakut25_NGC188SED}, despite both analyses being based on spectral energy distributions of member stars. The spread in the measured extinctions could indicate that there is differential reddening across the cluster. This was suggested in \citet{Cinar24_NGC188SED}, who reported larger $A_V$ measurements at lower RAs. 

\begin{figure}[ht!]
\plotone{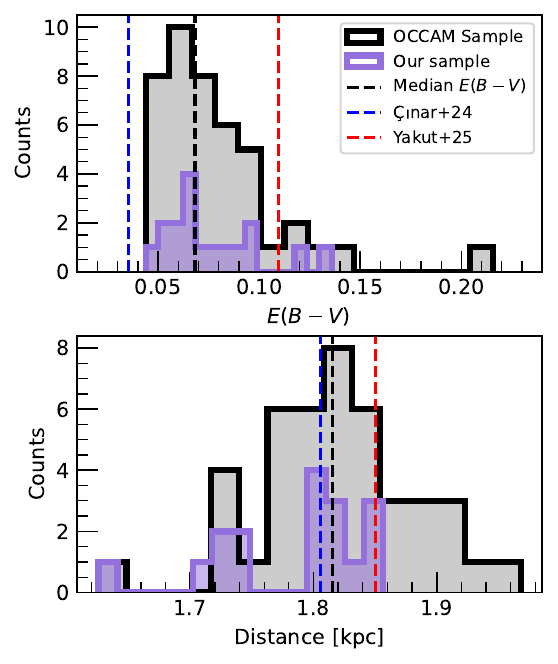}
\caption{\textbf{Top}: Distribution of the dust extinction, $E(B-V)$, for the OCCAM membership sample (black) and our sample (purple). The median extinction calculated from the OCCAM sample is illustrated by the black vertical line. We also demonstrate the measured $E(B-V)$ for NGC 188 from two previous studies by vertical dashed lines: \citet{Cinar24_NGC188SED} (blue) \& \citet{Yakut25_NGC188SED} (red). \textbf{Bottom:} Same for the distances from the \citet{BailerJones21_GaiaDistances} catalog.
\label{fig:ebv_distances}}
\end{figure}

Instead of adopting a single value for the extinction, we use the individual $E(B-V)$ values from \citet{Zhang23}, who infer line-of-sight dust extinctions from Gaia XP spectra. Stars not contained in this catalog were assigned the median extinction calculated from the OCCAM membership sample of $A_V=0.21$ or $E(B-V) = 0.07$  (see top panel of Fig.~\ref{fig:ebv_distances}) assuming a universal value for the dust extinction ratio of $R_V = 3.1$. Our inferred average cluster extinction falls between the two values reported in the aforementioned spectral energy distribution analyses. We emphasize that a detailed high-resolution survey of differential dust extinction is needed for this cluster.

We use the precise photometric magnitudes from Gaia DR3 \citep{GaiaDR3_releasepaper} to calculate a stellar radius (hereafter the Gaia radius) for each star. This approach first involves estimating a bolometric luminosity using:
\begin{equation}
    \log(L/L_{\odot}) = -0.4\left(M_G + BC_G - M_{\rm bol,\odot}\right)
\end{equation}
where $M_G$ is the absolute Gaia G-band magnitude, $BC_G$ is the bolometric correction in the G-band, and $M_{\rm bol,\odot}$ is the adopted Solar bolometric magnitude of 4.74 \citep{Mamajek15_mbol}. Bolometric corrections were computed using the \texttt{isochrones} \citep{Morton15_isochronepython} python package. This method estimates the bolometric corrections by interpolating over a grid of MIST isochrones \citep{Choi16_MIST}, given a star's $\Teff$, $\logg$ and metallicity. These quantities were taken from the spectroscopic surveys; either APOGEE DR17 catalog or the Gaia XP XGBoost catalog for stars not included in the former. 

The absolute Gaia magnitude was calculated from the apparent G-band magnitude ($G_{\rm mag}$), the distance ($r$), and the extinction in the G-band via: 
\begin{equation}
   M_G = G_{\rm mag} + 5 - 5\log_{10}(r) - A_{\rm G}
\end{equation}
Visual-band extinctions were converted to the Gaia $G$-band using extinction coefficients from \citet{Riello20_GaiaExtinction}\footnote{See description of method at \url{https://www.cosmos.esa.int/web/gaia/edr3-extinction-law}}. Distances were adopted from \citet{BailerJones21_GaiaDistances}, who infer probabilistic estimates from Gaia parallaxes using priors that incorporate Gaia colors, magnitudes, and extinction. Following \citet{Buder25_GalahDR4}, we use the photogeometric distances, where the distribution of distances for the cluster are shown in Figure~\ref{fig:ebv_distances}. Although \citet{BailerJones21_GaiaDistances} provide asymmetric uncertainties, we adopt the larger of the two bounds as the distance uncertainty to simplify the radius uncertainty propagation.

The Gaia radius is computed from the bolometric luminosity and $T_{\rm eff}$ via the Stefan–Boltzmann relation ($R\propto L^{1/2}\Teff^{-2}$). The resulting radii, along with the adopted temperatures and extinctions, are provided in Appendix~\ref{sec:A1_ParamTables}.


\section{Analysis of Seismic Masses for NGC 188}
\label{sec:mass_results}
\subsection{Seismic Mass Measurements}
\label{sec:calculating_masses}
We calculate the asteroseismic masses of red giants in NGC 188 using (following the notation of \citealt{Sharma2016_asfgridpaper}):
\begin{equation}
\label{eq:numax_mass_eq}
    \frac{M}{M_{\odot}}\simeq\left(\frac{\nu_{\text{max}}}{f_{\numax}\nu_{\text{max},\odot}}\right)\left(\frac{R}{R_{\odot}}\right)^2\left(\frac{T_{\text{eff}}}{T_{\text{eff},\odot}}\right)^{0.5}
\end{equation}
together with our measured $\numax$ (Sec.~\ref{sec:seismic_analysis}), Gaia radii and spectroscopic $\Teff$ (Sec.~\ref{sec:stellar_parameters}). We adopt $\nu_{\text{max},\odot} = 3090\pm30\:\mu$Hz \citep{Huber11_solar_syd_values} and $T_{\text{eff},\odot} = 5772\pm0.8~\mathrm{K}$ \citep{Mamajek15_B3}. The $f_{\numax}$ term is a correction factor that accounts for the observed discrepancy between the seismic and Gaia radii scales \citep{Sharma16_asfgrid}. By incorporating a $f_{\numax}$-correction into Eq.~\ref{eq:numax_mass_eq}, \citet{Ash25} found that the dispersion in the seismic masses for NGC 6791 stars is reduced, and the resulting masses are consistent with independent estimates from isochrones and eclipsing binaries. We adopt the $f_{\numax}$-corrections from the APOKASC-3 catalog using the Garstec+Mosser model\footnote{APOKASC-3 provides $f_{\numax}$-corrections for three different models. We tested each method and found no statistical differences between the calculated average masses. See more details about the $f_{\numax}$ correction analyses in Appendix~\ref{sec:A2_fnumax_corr}}. We note that the definition for the seismic mass scaling relation in the APOKASC-3 study has the $f_{\numax}$-correction as the reciprocal of our definition in Eq.~\ref{eq:numax_mass_eq}. Hence, we implemented the $\numax$-corrections as $f_{\numax}^{-1}$ into Eq.~\ref{eq:numax_mass_eq}.

\begin{figure*}[ht!]
\plotone{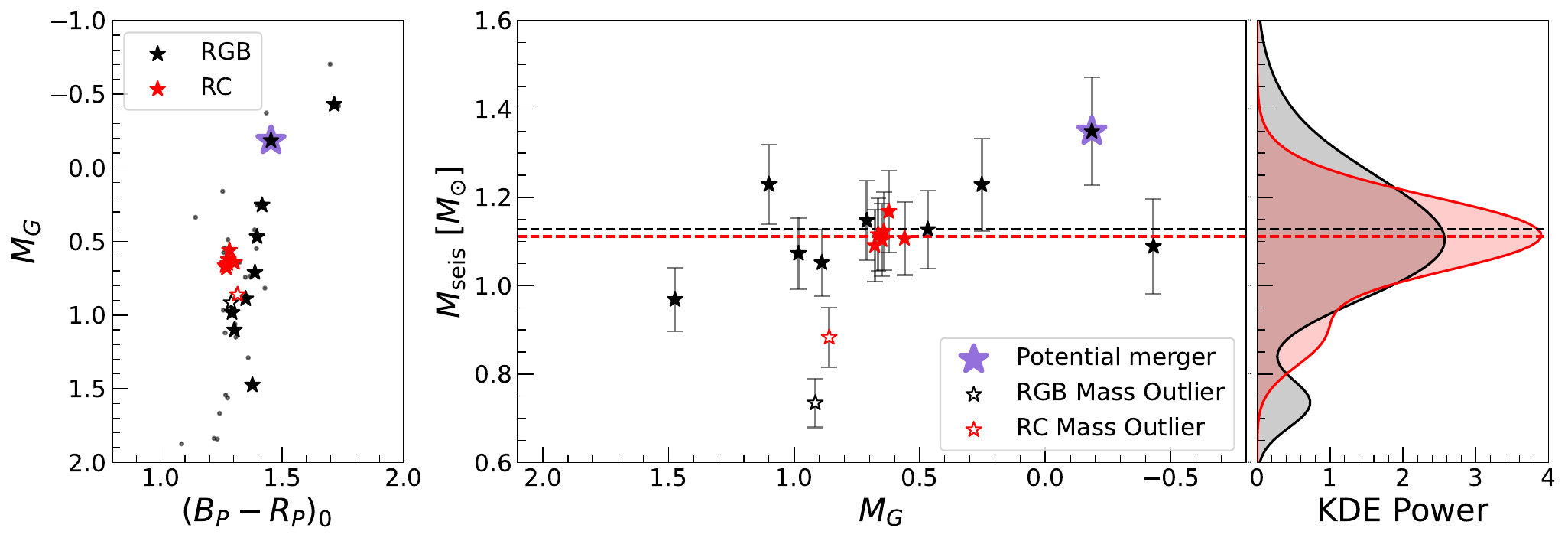}
\caption{\textbf{Left:} Absolute Gaia dust-corrected CMD for NGC 188. The asteroseismic sample is indicated by the star symbols, and separated into evolutionary phase (RGB in black and RC in red). \textbf{Middle:} Asteroseismic masses for our NGC 188 sample. Average masses are illustrated by the horizontal dashed lines for each evolutionary phase. Stars identified as mass outliers are distinguished by open symbols. Additionally, we have identified a star as a potential merger (Sec.~\ref{sec:potential_merger}), which is indicated by the purple border. \textbf{Right:} KDE of the asteroseismic masses demonstrating the distribution of seismic masses and their associated uncertainties. 
\label{fig:NGC188_masses}}
\end{figure*}

The resulting individual masses are shown in Figure~\ref{fig:NGC188_masses}, where the left panel displays absolute CMD, the middle panel shows the masses versus absolute Gaia magnitude, and the right panel shows the corresponding kernel density estimates (KDEs) computed using the individual random mass uncertainties following the method in \citet{Howell22_M4}.

We define mass outliers as stars whose masses deviate by more than three standard deviations from the median mass for their respective evolutionary phase. These stars are indicated by open symbols in Figure~\ref{fig:NGC188_masses}, and their presence contributes to the shoulder observed in the KDEs. Potential explanations for these outliers are discussed in Section~\ref{sec:mass_outliers}. In the same figure, we also highlight the most massive star in our sample (TIC 461601491) with a purple border. Although it is not classified as a mass outlier, its mass is substantially higher than the average RGB mass. Possible evolutionary scenarios for this star are explored in Section~\ref{sec:potential_merger}.

Excluding the two mass outliers, we calculate the median mass for each evolutionary phase as $M_{\text{RGB}} = 1.13\pm0.04~\msun$ and $M_{\text{RC}} = 1.11\pm0.01~\msun$, which are denoted by dashed horizontal lines in Figure~\ref{fig:NGC188_masses}. The uncertainties on the average masses are calculated as the standard error on the mean, which is a measure of the scatter around the average mass. Comparing our median RGB seismic masses to the independent estimates from a red giant star in an eclipsing binary system ($M = 1.174\pm0.012~\msun$; \citealt{Yakut25_NGC188SED}) and a PARSEC isochrone at the RGB bump ($M = 1.18~\msun$; Table~\ref{tab:isochrones}), we found agreement within $2\sigma$ uncertainties. We discuss inherent systematics in the absolute average mass estimates in Section~\ref{sec:systematic_uncertainties}.

A direct comparison between the fractional uncertainties on the median masses in our study and those from seismically studied open clusters with \textit{Kepler} can be used to gauge the efficacy of using \textit{TESS} photometry to measure cluster masses. By dividing the average mass uncertainty by the average mass, we calculate a fractional average mass uncertainty of 3.5\% for our RGB sample and 0.9\% for our RC sample. \citet{Ash25} reported fractional uncertainties of the median masses of 0.9-1.2\% for the \textit{Kepler} clusters NGC 6971 and NGC 6819. Additionally, our fractional cluster mass uncertainties are comparable to the reported median fractional uncertainty on the seismic mass for the APOKASC-3 gold sample of 3.8\%. \textit{This demonstrates that by doing a boutique analysis of the \textit{TESS} photometry, we can measure the average cluster masses to a similar precision to previous \textit{Kepler} studies of clusters and field stars.} 



We estimate the integrated mass loss on the RGB ($\Delta M$) as the difference between the median RGB and RC masses, finding $\Delta M = 0.02 \pm 0.04~M_\odot$. The uncertainty is calculated  by adding in quadrature the median mass uncertainties. Our integrated mass loss measurement also incorporates the difference in the birth mass of stars in the RGB and RC evolutionary phases, where our best fit PARSEC isochrone (Table~\ref{tab:isochrones}) predicts that the RC population is initially $0.01~\msun$ more massive than the RGB bump population. Assuming an evolutionary initial mass difference of $0.01~\msun$, our measured mass loss would be $0.03~\msun$, which is smaller than the predicted mass loss from our PARSEC isochrone of $0.07~\msun$. This could suggest that the adopted Reimers' scaling parameter of $\eta_R = 0.1$ in the isochrone is too large to describe the mass loss for stars in this cluster. Another estimate for $\eta_R$ can be derived from the metallicity and mass-dependent prescription provided in Eq.~2 of \citet{Roberts25}, which predicts of a similar value of $\eta_R = 0.11$. Our result contributes to growing skepticism in the literature of the accuracy of mass loss schemes, such as Reimers', to calculate the mass loss rate in stellar models for low mass stars. This is emphasized by the tension between mass loss-metallicity trends between the metal-poor and metal-rich populations, where we see a stark dichotomy in these relations that is not predicted from current mass loss schemes \citep{Li25_massloss}. We note that our small integrated mass loss for NGC 188 is consistent with the measured mass loss-metallicity trend in \citet{Li25_massloss} at approximately solar-metallicity.

\subsection{Quantifying the Systematic Uncertainties on the Average Masses \& Mass Loss}
\label{sec:systematic_uncertainties}

To assess the systematic effect on the average masses, we consider three parameters where the individual uncertainties do not reflect the systematic differences reported in the literature. As detailed in Section~\ref{sec:stellar_parameters}, recent measurements for the extinction of NGC 188 span the range $A_V = 0.11-0.34$ \citep{Cinar24_NGC188SED,Yakut25_NGC188SED}, where we adopt a value approximately central in this range of $A_V = 0.21$ (Fig.~\ref{fig:ebv_distances}). Similarly there have been inconsistent distance measurements reported for this cluster: $1850\pm12$~pc in \citet{Yakut25_NGC188SED}, and $1806\pm21$~pc in \citet{Cinar24_NGC188SED}, where the latter distance is similar to the median value used in this study (Fig.~\ref{fig:ebv_distances}). Additionally, offsets in $\Teff$ scales have been reported in different spectroscopic surveys and data releases. For example, we use spectroscopic $\Teff$ measurements from APOGEE DR17, however there is a median offset of $60$~K when comparing to APOGEE DR19 temperatures for our NGC 188 sample \citep{APOGEEDR19,Meszaros25_ASPCAP}. We test whether adopting the different literature values for extinction and distance, and the effect of differing $\Teff$ scales cause significant discrepancies in the average RGB and RC cluster masses. We also test the effect on the integrated mass loss, which is calculated as differences between the average masses of the two evolutionary phases\footnote{We note that there are known dispersions in the measurement of the asteroseismic quantities between pipelines that would lead to systematics in the derived seismic masses. We do not investigate this here as we only use one asteroseismic pipeline. See  \citet{Pinsonneault18_APOKASC2,Pinsonneault2025_APOKASC3} for more details}.

Table~\ref{tab:Mass_systematics_tab} reports the differences between our average masses and mass loss, and the average masses derived after propagating through the new parameter values. Both the \citet{Yakut25_NGC188SED} and \citet{Cinar24_NGC188SED} studies use spectral energy distribution fits to infer the line-of-sight extinction and distance; as such these two parameters are correlated. Therefore, we investigate the systematic effect to the seismic masses and mass loss when propagating these parameters simultaneously. We also provide the difference in the masses when propagating the parameters individually, however we do not consider them further in our analysis of the systematics. 

We quantify the effect of the systematics on the average RGB and RC mass, and the integrated mass loss by adding in quadrature the systematic difference in mass for the correlated distance and extinction value, and the $\Teff$ scale value. We find a cluster average RGB mass of $M_{\text{RGB}} = 1.13\pm0.04$(random)$^{+0.12}_{-0.19}$(sys)$~\msun$ and RC mass of $M_{\text{RC}} = 1.11\pm0.01$(random)$^{+0.11}_{-0.19}$(sys)$~\msun$. These systematic uncertainties highlight the importance of the adopted cluster parameters and $\Teff$ scales in determining the absolute cluster masses, as they are significantly larger compared to the intrinsic scatter in the individual masses represented by the random uncertainty. However, the systematics are less significant for the difference between the evolutionary masses, where our derived integrated mass loss is $\Delta M = 0.02 \pm 0.04$(random)$\pm0.01$(sys)$\msun$.  

\begin{deluxetable*}{lccccc}
\digitalasset
\tablewidth{0pt}
\tablecaption{Differences in the median cluster masses for the RGB ($\delta M _{\text{RGB}}$) and RC ($\delta M _{\text{RC}}$) when adopting different recent literature values for the dust extinction ($A_V$) and distance for NGC 188. These differences are calculated as our reported value minus the value derived from propagating in the systematics. We also investigate the effect of different $\Teff$ scales, where we use the median offset between APOGEE DR17 and APOGEE DR19 catalogs for our sample. The last column reports the differences between our nominal integrated RGB mass loss and the calculated mass loss using the associated literature values ($\delta ( M _{\text{RGB}} -  M _{\text{RC}})$). \label{tab:Mass_systematics_tab}}
\tablehead{
\colhead{Parameter} &  \colhead{Reference} & \colhead{Value} & \colhead{$\delta M _{\text{RGB}}$ ($M_{\odot}$)} & \colhead{$\delta M _{\text{RC}}$ ($M_{\odot}$)} & \colhead{$\delta ( M _{\text{RGB}} -  M _{\text{RC}})$ ($M_{\odot}$)} }
\startdata
$A_V$  & \citealt{Cinar24_NGC188SED} & 0.11~mag & 0.15 & 0.06 & 0.09  \\
$A_V$  & \citealt{Yakut25_NGC188SED} & 0.34~mag & -0.03 & -0.13 & 0.1\\
Distance & \citealt{Cinar24_NGC188SED} & 1806~pc  & 0 & 0 & 0 \\
Distance & \citealt{Yakut25_NGC188SED} & 1850~pc  & -0.06 & -0.06 & 0 \\
Distance \& $A_V$ & \citealt{Cinar24_NGC188SED}& -  & 0.09 & 0.08 & 0.01 \\
Distance \& $A_V$ & \citealt{Yakut25_NGC188SED}& -  & -0.17 & -0.17 & 0.01 \\
$\Teff$ scale  & APOGEE DR17 vs DR19 &  $\pm60$~K & $\pm0.08$ & $\pm0.08$ & $\pm0.01$ \\
\enddata
\end{deluxetable*}

\subsection{Relationship between Chemical Abundances \& Seismic Masses}
\subsubsection{[C/N] Abundances}
\label{sec:CN_abundances}

The populations of open clusters are expected to have homogeneous chemical signatures. Therefore, evidence of significant differences in surface abundances between stars at similar evolution points could be indicative of binary interactions. During first dredge up, RGB stars mix hydrogen-burning products from the core to the surface, where the subsequent [C/N] abundance is dependent on mass and metallicity \citep{Iben1965}. Hence, for the near mono-mass and metallicity red giant populations in open clusters, we expect all RGB stars after first dredge up and RC stars to have the same [C/N] abundance. Any significant deviations in [C/N] could indicate the occurrence of a non-standard evolutionary event; such as mass transfer to/from a binary companion or evidence of a past merger. 

\begin{figure}[ht!]
\plotone{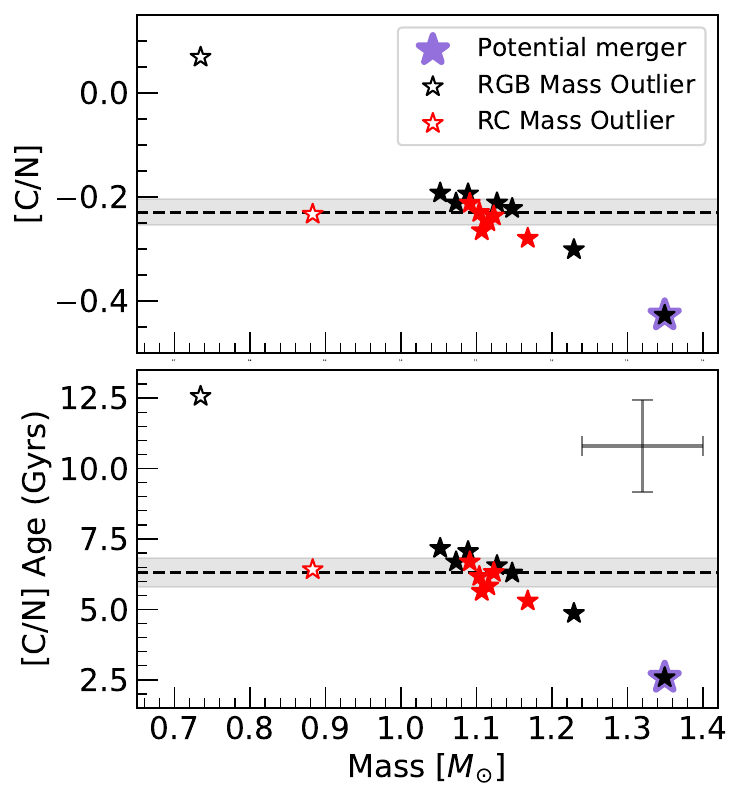}
\caption{\textbf{Top:} [C/N] abundances against seismic mass. Colour coding and symbols are the same as Fig.~\ref{fig:NGC188_masses}. The median [C/N] abundance and standard error on the mean of $-0.23 \pm 0.03$ is indicated by the dashed horizontal line and shaded region.  \textbf{Bottom:} Same as the top panel, but for the [C/N]-ages on the vertical axis. The median [C/N]-age value is reported in Table~\ref{tab:NGC188_age_estimates}, and demonstrated here by the dashed horizontal line. The standard error on the mean of 0.5~Gyrs is illustrated by the shaded region, and a typical error bar is shown in the top right. 
\label{fig:CN_ages_masses}}
\end{figure}

For our NGC 188 sample, we used [C/N] abundances from the APOGEE DR17 catalog and measured a median [C/N] abundance of $-0.23\pm0.03$. We show the relationship between the [C/N] abundances and seismic masses in Figure~\ref{fig:CN_ages_masses}, where we identify two outliers with [C/N] that deviate significantly from the average abundance. Interestingly, the RGB star that is enhanced in [C/N] has been identified as a mass outlier (indicated with the open symbol) and the RGB star depleted in [C/N] is speculated to be a potential merger remnant. We discuss these stars further in Section~\ref{sec:mass_outliers} and \ref{sec:potential_merger}, respectively. 

\subsubsection{Lithium Abundances}
The formation of lithium-rich giants remains an open question in stellar physics, due to the ease in destroying this element during the RGB first dredge up event. One theory for this lithium enrichment is related to binary interactions. It has been proposed that the formation mechanism for Li-rich giants is mass transfer from a Li-rich asymptotic giant branch star \citep{Sackmann1992_LirichAGB,AguileraGomez16, Sayeed25} to a main sequence star, where we detect the latter when it has evolved into a giant. Recent evidence from \citet{Sayeed24} has shown that Li-rich stars at the base of the RGB on average have a higher abundance of mean \textit{s}-process elements compared to Li-normal stars, indicative of mass transfer from an asymptotic giant branch companion. They also found that Li-rich giants are more likely to be fast rotators, further supporting a binary mass transfer scenario. Using our seismic mass estimates, we investigated whether there is a correlation between anomalous masses and Li-enrichment in our NGC 188 sample.

\begin{figure}[ht!]
\plotone{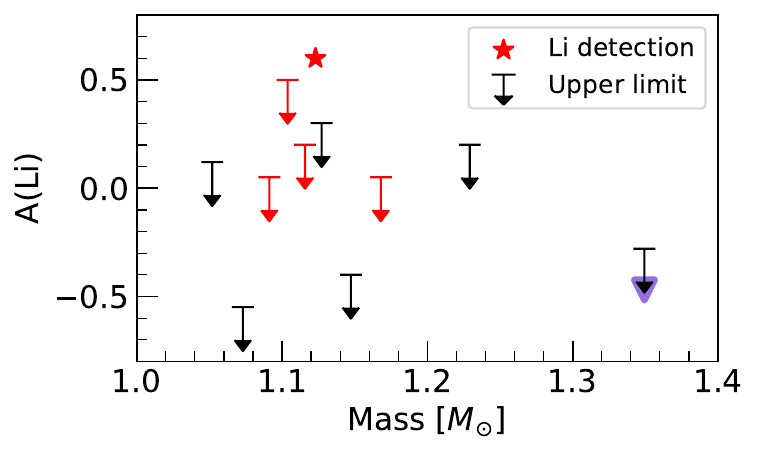}
\caption{Lithium abundances plotted against seismic masses. Color coding is the same as Fig.~\ref{fig:NGC188_masses}. Upper limits (arrows) are distinguished from lithium detections (star symbol). 
\label{fig:Li_masses}}
\end{figure}

Open clusters are commonly used to investigate the origins of the anomalous Li-rich giants because their RGB masses and ages are well determined, especially when combined with asteroseismology \citep[e.g.][]{Carlberg15}. Lithium in NGC 188 giants was measured in the WIYN open cluster study \citep[WOCS; ][]{Sun2022_NGC188Li}, where there is an overlap of 11 giants with our seismic sample. We present the Li measurements plotted against seismic mass in Figure~\ref{fig:Li_masses}. One RC star from our sample was identified as a Li-enhanced giant, TIC 461620563 (WOCS ID 6353). This star has a seismic mass of $1.12\pm0.09~\msun$, which is consistent with our average RC mass within the uncertainty. Furthermore, it was reported as a single star with low radial velocity variability in the WOCS survey \citep{Narayan25_WOCSRV}. Hence, we suggest this star has not undergone a mass transfer event due to a binary interaction, and a different process has led to the lithium enrichment. We note that \citet{Narayan25_WOCSRV} classified this star as belonging to the asymptotic giant branch evolutionary phase, however our mass estimate and de-reddened photometry suggest that this star most likely belongs to the RC evolutionary phase (Fig.~\ref{fig:NGC188_masses}), as assigned by this study. 

Our potential merger candidate (TIC 461601491 or WOCS 4843) was found to be in an eclipsing binary system in \citet{Narayan25_WOCSRV} with a period of 1241.7 days. However, it has a Li-upper limit that is similar to the other `Li-normal' cluster members. We discuss the results for this star further in \ref{sec:potential_merger}. Unfortunately, neither of our under-massive targets has Li measurements in the WOCS survey.

\subsection{Stars with Outlying Masses}
\label{sec:mass_outliers}
In Figure~\ref{fig:NGC188_masses}, we identified two statistical outliers with masses lower than the cluster average: the RGB star TIC 461593694 and the RC star TIC 461626065. Both stars were confirmed as cluster members via Gaia astrometry and APOGEE spectroscopy. Therefore, we instead consider that the outlying masses are a result of non-standard evolutionary events, where we propose that these stars are post-mass transfer remnants due to binary interactions. Initially, we checked for any indications of their binarity from two parameters: i) the VSCATTER parameter from the APOGEE DR17 catalog, which measures the observational scatter in the radial velocities from multiple APOGEE visits and could imply binarity if VSCATTER $>1$ km/s, and ii) the renormalized unit weight-error (RUWE) in the Gaia DR3 catalog, which describes the astrometric deviations from a single-source solution and can suggest the presence of a close binary companion if RUWE $>1.4$. Both stars had a VSCATTER $<0.06$~km/s, and a RUWE $<1.4$ (TIC 461626065 is close to the boundary with RUWE $=1.365$). Although these metrics are inconsistent with these stars being in binary systems, we cannot rule this theory out. We note that neither of these stars were included in the time-series radial velocity WOCS study of this cluster. Therefore, detailed spectroscopic follow up of these stars to search for radial velocity variations could reveal a binary companion, and support that these masses are caused by a stripping event. Nonetheless, we explore each star in more detail below.

\subsubsection{TIC 461593694}
The RGB star, TIC 461593694, is the least massive star in our sample with a mass of $M=0.73\pm0.06~\msun$. In the extinction-corrected CMD (left panel of Fig.~\ref{fig:NGC188_masses}), this star is not photometrically distinct from the general RGB sample. Comparing to other RGB stars at a similar evolutionary point, the star has a similar effective temperature and Gaia radius (see summary of results in Table~\ref{tab:NGC188_table}). However, its $\numax$ is lower, resulting in the lower mass. This star was categorized into our `gold' asteroseismic sample (Sec.~\ref{sec:gold_sample}), and as such we rule out the possibility of inaccuracies when measuring $\numax$ given the presence of a clear oscillation pattern. 

This star has a [C/N] abundance that is significantly larger than the cluster average (Fig.~\ref{fig:CN_ages_masses}). The [C/N] abundance on the RGB is a tracer of the mass prior to the first dredge up. A large difference in this abundance could suggest that a non-standard evolutionary event occurred before the first dredge up, e.g. a mass transfer event where the star has lost mass from its envelope. There have been detections of under-luminous sub-giant stars in open clusters, which are thought to have experienced enhanced mass loss \citep[e.g.][]{Leiner2017}. Referred to as sub-subgiants, one possible mechanism for this mass loss is that these objects are the remnants of a stripping event potentially due to a binary interaction. We speculate that these sub-subgiant stars could be the progenitors for our under-massive RGB star. This scenario supports the [C/N] measurement, as the mass stripping event would have occurred prior to the RGB phase. Consequently, this star would have a lower mass during the first dredge up, which culminates in a higher [C/N] abundance after envelope mixing \citep[e.g.][]{Roberts24_CN}. Further analyses is needed to determine the plausibility of this mass transfer scenario. 



\subsubsection{TIC 461626065}
The RC star, TIC 461626065, has a mass of $M=0.88\pm0.07~\msun$ and photometrically is fainter than the RC population in Figure~\ref{fig:NGC188_masses}. The star has a [C/N] abundance that agrees with the cluster average, suggesting that mass transfer could have occurred after the first dredge up while the star was ascending the RGB.

Post-mass transfer RC stars have been investigated with asteroseismology in the literature. \citet{Li2022_RCundermassive} discovered 32 RC stars in the \textit{Kepler} field with inferred ages that exceed the age of the Universe. This is a strong indication that this sample of stars were initially more massive and have since experienced a stripping event from a companion. Under-massive RC stars have also been discovered in seismically-studied open clusters. For example, \citet{Brogaard6791_undermassivestar} discovered an under-massive star in NGC 6791, and speculated that the star had undergone larger mass loss compared to other cluster members. Additionally, an under-massive star was detected in NGC 6819, where it was initially flagged as a seismic outlier (and reported as a non-member under the assumption of single-star evolution) in \citet{Stello2011_clustermembersfromseismology}, and was reclassified as an under-massive cluster member in \citet{Handberg2017}. That star is also Li-rich, where the lithium enrichment is thought to be correlated to the mass transfer process. A theoretical study was conducted by \citet{Matteuzzi24} for the NGC 6819 star using binary population models. Interestingly, they concluded that the likely scenario was that common-envelope mass transfer occurred during the RGB phase, where the low-mass main sequence companion does not survive and approximately $1~\msun$ of material is ejected into the intra-cluster medium. This theory could also describe the formation of our NGC 188 under-massive RC star, and is consistent with the [C/N] abundance result and the current lack of evidence of binarity. However, the feedback to the intra-cluster medium would be smaller for our under-massive RC star, where there is a difference of $\sim0.23~\msun$ between TIC 461626065 and the median RC mass for NGC 188. As emphasized by \citet{Matteuzzi24}, further evolutionary models are needed to fully understand these systems. 

Asteroseismic mass surveys of clusters are revealing stars that have experienced non-standard evolutionary pathways. The identification of these objects is important for multiple star system studies, and can be used to learn about binary interaction processes. Additionally, it would be interesting to measure the lithium abundance of TIC 461626065 to further investigate if there is a correlation between lithium enrichment and mass transfer between binaries.

\section{Age Results for NGC 188}
\label{sec:age_results}

NGC 188 has been the target of numerous age determination studies using a variety of techniques. We summarize these studies in Table~\ref{tab:NGC188_age_estimates}, which presents a spread of 2~Gyr among the reported ages. The precisions on these ages often do not reflect the inconsistency between the methods. In this study, we estimate stellar ages and their associated uncertainties using two approaches: (i) inferring seismic ages from isochrone fitting to the seismic masses and stellar parameters, and (ii) applying the [C/N]–age relation from \citet{Roberts25}, which derives ages from measured [C/N] abundances and metallicity.

\begin{deluxetable*}{lcc}
\digitalasset
\tablewidth{0pt}
\tablecaption{A summary of literature age estimates for NGC 188 from different methods. Our final seismic and [C/N] cluster age are provided at the bottom. \label{tab:NGC188_age_estimates}}
\tablehead{
\colhead{Age Method} & \colhead{Age (Gyrs)} & \colhead{Reference}} 
\startdata 
CMD Isochrone Fits & $7.0\pm0.1$ & \citealt{Bonatto05_NGC188isochrones} \\
... & $7.5\pm0.7$ & \citealt{Fornal07_NGC188isochrone} \\
... &  $5.78\pm0.03$ \& $6.45\pm0.04$ & \citealt{Hills15_NGC188isochrones}\\
... & $7.08\pm1.58$ & \citealt{Cantat-Gaudin2020_ocmembership} \\
... & $7.65\pm1.00$ & \citealt{Cinar24_NGC188SED} \\
White Dwarf Cooling Track & $\geq 5$ & \citealt{Andreuzzi02_WDAge} \\
Eclipsing Binaries & $6.2\pm0.2$ & \citealt{Meibom09_EBage} \\
... & $6.41\pm0.28$ & \citealt{Yakut25_NGC188SED} \\
Spectral Energy Distributions & $7.78\pm0.23$ & \citealt{Cinar24_NGC188SED} \\
\hline
Our Asteroseismic Age & $7.0\pm0.9$ &  Sec.~\ref{sec:seismic_ages} \\
$[\text{C/N}]$-Age Relation & $6.3\pm0.5$ & Sec.~\ref{sec:cn_ages}; using relation in \citealt{Roberts25}\\
\enddata
\end{deluxetable*}

\subsection{Seismic Ages}
\label{sec:seismic_ages}
We follow the method outlined in \citet{Morales2025} of inferring ages from seismic masses and stellar parameters. In summary, they employ the \texttt{kiauhoku} \citep{Claytor2020_kiahoku} stellar model grid interpolator to determine the best fitting evolutionary tracks to the input priors. They use four different stellar model grids: MIST \citep{Choi16_MIST}, DSEP \citep{Chaboyer2001_DSEP1,Dotter2008_DSEP}, GARSTEC \citep{Weiss2008_GARSTEC}, and YREC \citep{Pinsonneault1989_YREC}. A Monte-Carlo sampling procedure was used to estimate an age uncertainty for individual stars \citep{Tayar22_griduncertainties}. Due to the uncertainty in mass loss rates in stellar models, we only infer seismic ages for our RGB sample as it does not rely on an uncalibrated Reimers' $\eta_R$ efficiency. 

As demonstrated in numerous studies \citep[e.g.][]{SilvaAguirre2020_aarhusredgiantchallenge,Li24_stellarmodellingstudy,Morales2025}, the differences between stellar models and input physics can result in significant systematics between the derived seismic ages, where \citet{Morales2025} measured a mean offset of 10\% between model grids when including seismic constraints. We further emphasize this discrepancy in Table~\ref{tab:age_systematics}, which provides the median cluster age and the associated standard error on the mean calculated from the different model grids. Mass outliers identified in Section~\ref{sec:calculating_masses} were excluded from the median age estimates to reduce potential bias. The median cluster ages obtained from the DSEP, GARSTEC, and YREC models are mutually consistent, but they only weakly agree with literature age estimates for NGC 188 (Table~\ref{tab:NGC188_age_estimates}). In contrast, the median age derived from the MIST models of $7.0 \pm 0.9$~Gyr, is in better agreement with literature age measurements, and provides the smallest scatter between the ages of cluster members (represented by the smallest precision of all the models). Hence, we adopt the MIST-based ages as our final seismic age estimates.


Our cluster age uncertainty, calculated as the standard error on the mean, is larger than many absolute age uncertainties reported in the literature. We believe that our measurement represents a realistic precision on the cluster age given it directly corresponds to the observed scatter in the individual stellar ages. Inspection of the MIST-derived ages (Fig.~\ref{fig:AgeComp} and Table~\ref{tab:NGC188_table}) demonstrates the range of 6~Gyrs between the individual cluster members. A similar large spread was also reported by \citet{Tayar25}, who found that the scatter between the ages of stellar cluster members was significantly larger than absolute age uncertainties reported in previous studies and larger than the discrepant ages from different model grids. These results highlight the need for continued improvement in seismic age inference methods. Star clusters provide critical benchmarks for calibrating these techniques, which can then be applied to robustly determine ages of field red giants. Achieving accurate seismic ages is essential for Galactic archaeology studies \citep[e.g.][]{Pinsonneault2025_APOKASC3}.


\begin{deluxetable*}{lc}
\digitalasset
\tablewidth{0pt}
\tablecaption{Inferred median cluster ages from different model grids. The age uncertainty is calculated as the standard error on the mean. \label{tab:age_systematics}}
\tablehead{
\colhead{Model Grid} & \colhead{Median Cluster Age [Gyrs]}}
\startdata
MIST & $7.0\pm0.9$ \\
DSEP & $8.1\pm1.0$ \\
GARSTEC & $8.1\pm1.0$ \\
YREC & $8.3\pm1.0$ \\
\enddata
\end{deluxetable*}

\subsection{[C/N] Ages}
\label{sec:cn_ages}
We obtain independent age estimates for individual stars using a [C/N]–age relation. This chemical-clock method relies on the fact that during first dredge-up on the RGB, stars mix processed material from the core to the surface in a manner that depends on stellar mass and metallicity. Hence, [C/N] and metallicity can be used as a diagnostic for birth mass and the age for RGB and RC stars after first dredge up. We use Equation~4 from \citet{Roberts25}, and [C/N] and [Fe/H] abundances from the APOGEE DR17 survey to estimate the [C/N]-ages for our NGC 188 sample (the two stars not contained within the APOGEE DR17 survey were excluded from this analysis). We demonstrate the resulting ages in Figure~\ref{fig:CN_ages_masses}. Individual age uncertainties are adopted as the mean uncertainty of 1.64~Gyr from \citet{Roberts25}. We derive a median cluster [C/N] age and standard error on the mean of $6.3\pm0.5$~Gyrs, which is consistent with the literature ages presented in Table~\ref{tab:NGC188_age_estimates}.

\begin{figure}[ht!]
\plotone{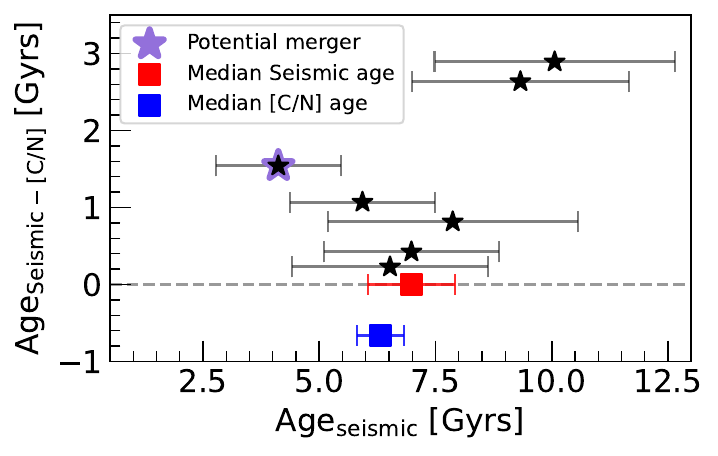}
\caption{Differences between our calculated seismic ages to the [C/N] ages for the overlapping NGC 188 RGB sample (black stars). The dashed grey line indicates the zero-point offset. The median seismic (red) and [C/N] (blue) cluster age shown with square symbols. The RGB mass outlier identified in Section~\ref{sec:mass_outliers} is not included, as it had a unrealistic residual age of $+23$~Gyrs. 
\label{fig:AgeComp}}
\end{figure}

The [C/N]-age relation in \citet{Roberts25} was calibrated on APOKASC-3 seismic ages and APOGEE DR17 abundances, and thus the ages are on a similar scale to our seismic ages. In Fig~\ref{fig:AgeComp}, we directly compare the [C/N] ages to our inferred seismic ages for the overlapping RGB sample. The seismic ages are consistently larger than the [C/N] ages, with discrepancies up to 3~Gyrs for the oldest stars in our sample. Despite the larger variance between the individual masses, there is agreement within $1\sigma$ uncertainties in the median cluster age from both methods. 

\section{Extended Asteroseismic Analysis of the Gold Sample in NGC 188}
\label{sec:gold_sample}

\subsection{Measurement of $\Delta\nu$}

We identified a sub-sample of 9 stars in NGC 188 with clear oscillation modes. For this `gold' sample of stars, we extended the asteroseismic analysis to include the measurement of the large frequency spacing, $\Delta\nu$. We used the \texttt{pySYD} pipeline \citep{Chontos21_pySYD}, where $\Delta\nu$ is estimated from an auto-correlation function that searches for regular frequency spacings between peaks in the oscillation envelope. The accuracy of this method is dependent on a clear oscillation pattern to measure $\Delta\nu$, and is not optimized for finding regular spacings when there is a lower number of radial orders (e.g. at low $\numax$) or when there is increased noise. This is why we have limited the measurement of $\Delta\nu$ to our gold sample. We also tested whether the estimated $\numax$ from the two asteroseismic pipeline \texttt{pyMON} and \texttt{pySYD} were consistent, and found that the two measurements agreed within 1\% (except for the two lowest $\numax$ stars which were consistent within 3\%).

It is well established that there are discrepancies between the theoretical $\Delta\nu\propto\overline{\rho}^{1/2}$ relation and the observed frequency pattern  \citep{Stello09_dnu_numax_relation,White11_dnucorr1,Miglio2012_OCs,Miglio13_dnucorr3}. To correct for these deviations, a factor $f_{\Delta\nu}$ is introduced into the seismic scaling relations. This correction factor is dependent on metallicity, initial mass, and evolution stage, and is calculated by interpolating over a grid of models. We implement the \texttt{Asfgrid} v0.0.6 \citep{Sharma2016_asfgridpaper,Stello2022_asfgrid_v2} to determine $f_{\Delta\nu}$ for our gold sample.

We calculate the $\Delta\nu$-dependent seismic masses for our gold sample using our estimates for $\numax$, $f_{\numax}$, $\Delta\nu$ and $\Teff$, and the equation:
\begin{equation}
\label{eq:numax_dnu_mass_eq}
    \frac{M}{M_{\odot}}\simeq\left(\frac{\nu_{\text{max}}}{f_{\numax}\nu_{\text{max},\odot}}\right)^3 \left(\frac{\Delta\nu}{f_{\Delta\nu}\Delta\nu_{\odot}}\right)^{-4}\left(\frac{T_{\text{eff}}}{T_{\text{eff},\odot}}\right)^{1.5}
\end{equation}
where we used a solar large frequency spacing of $\Delta\nu_{\odot} = 135.1\pm0.1$ \citep{Huber11_solar_syd_values}. To differentiate the seismic mass scales, we refer to the masses calculated using Eq.~\ref{eq:numax_dnu_mass_eq} as $M_{\Delta\nu,\numax}$; whereas the nominal masses calculated via Eq.~\ref{eq:numax_mass_eq} are referred to as $M_{\numax}$. The $M_{\Delta\nu,\numax}$ values are provided in Table~\ref{tab:gold_sample}, along with our estimated $\Delta\nu$ and $f_{\Delta\nu}$. We use the same values for $f_{\numax}$ as in Section~\ref{sec:mass_results}, where again the $\numax$-corrections adopted from APOKASC-3 are implemented into Eq.~\ref{eq:numax_dnu_mass_eq} as $f_{\numax}^{-1}$. 

\begin{figure}[ht!]
\plotone{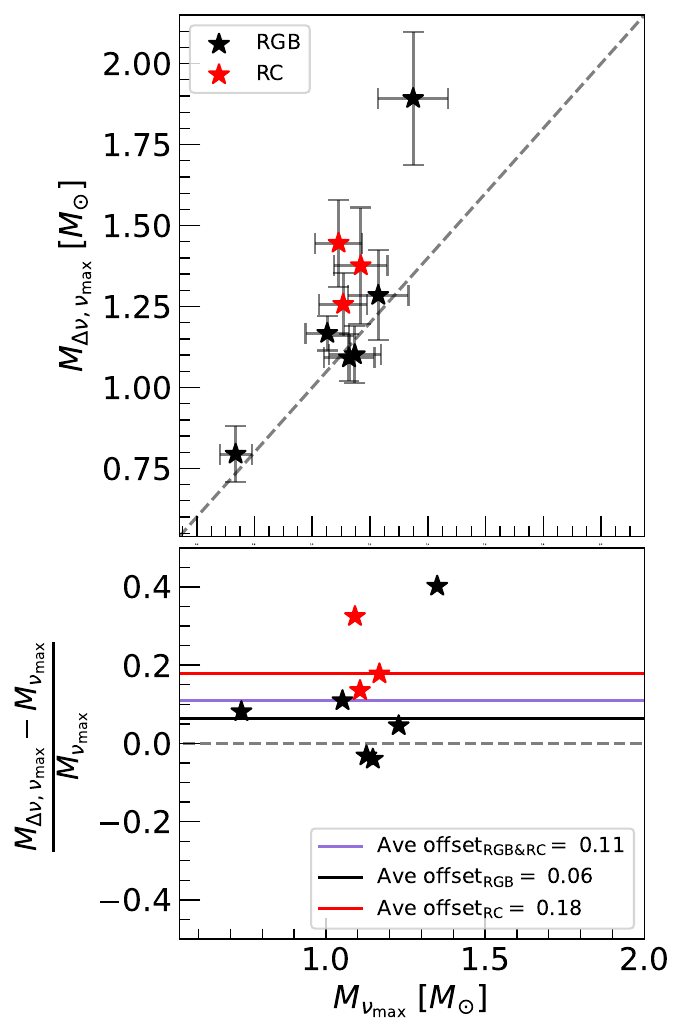}
\caption{\textbf{Top:} Comparison between the $\Delta\nu$-dependent seismic masses ($M_{\Delta\nu,\numax}$) and the seismic mass calculated from $\numax$ and a Gaia radius ($M_{\numax}$). Dashed grey line indicates a unity relationship. \textbf{Bottom:} The fractional residuals between the two seismic mass scales. The dashed line indicates the zero-point. The median fractional offset between the two mass scales for both the RGB and RC is indicated by the horizontal purple line, the RGB sample by the black line, and the RC sample by the red line.
\label{fig:MassComp}}
\end{figure}

Previous asteroseismic analyses of clusters have shown discrepancies between $M_{\Delta\nu,\numax}$ and $M_{\numax}$ \cite[e.g.][]{Miglio2012_OCs,Howell22_M4, Howell24_M80, Reyes25_M67}. Figure~\ref{fig:MassComp} shows a comparison between the two mass scales for our gold sample, where we found a median fractional offset of 11\%. Restricting the sample to the RC population, the average fractional offset increases to 18\%. This could suggest that the $f_{\Delta\nu}$ factor is not properly correcting the masses of our RC stars, where $f_{\Delta\nu}=1$ was derived for the entire RC sample (Table~\ref{tab:gold_sample}). We also note that our RC sample consists of three stars, one of which has a fractional offset of 32\% that is biasing the median fractional offset to be larger for this evolutionary phase. As such, the observed RC discrepancy may be driven by small number statistics.

Excluding the RC stars reduces the average fractional offset in masses for the RGB stars to 6\%, which is still larger than values reported from \textit{Kepler} data \citep[e.g.][]{Reyes25_M67}. This discrepancy may arise from less accurate $\Delta\nu$ measurements, likely reflecting the lower quality of the \textit{TESS} photometry compared to \textit{Kepler}. We emphasise that the $M_{\numax}$ values have been shown to be more robust for cluster seismic analyses, and as such are adopted as the final masses for our study \citep[e.g.][]{Miglio2012_OCs,Howell22_M4,Reyes25_M67}.

\begin{deluxetable*}{lccccc}
\digitalasset
\tablewidth{0pt}
\tablecaption{Measured $\Delta\nu$, $f_{\Delta\nu}$ corrections, masses using both seismic quantities ($M_{\Delta\nu,\numax}$; Eq.~\ref{eq:numax_dnu_mass_eq}) \& dervied dipole-mode visibilities for the NGC 188 gold sample. TIC IDs and evolutionary labels are also provided. \label{tab:gold_sample}}
\tablehead{
\colhead{TIC ID} & Evol & \colhead{$\Dnu$ $(\mu \mathrm{Hz})$} & \colhead{$f_{\Dnu}$} & \colhead{$M_{\Delta\nu,\numax}$ ($M_{\odot}$)} & \colhead{$V^2_{\ell=1}$}}
\startdata
461595568 & RGB & $3.83\pm0.03$ & $0.97$ & $1.17\pm0.05$ & $1.57\pm0.40$ \\
461593694 & RGB & $3.64\pm0.09$ & $0.97$ & $0.79\pm0.09$ & $2.68\pm0.67$ \\
461599427 & RC & $3.70\pm0.06$ & $1$ & $1.26\pm0.10$ & $5.40\pm0.70$ \\
461601138 & RGB & $2.45\pm0.05$ & $0.99$ & $1.28\pm0.14$ & $1.10\pm0.44$ \\
461601229 & RGB & $2.86\pm0.04$ & $0.97$ & $1.09\pm0.07$ & $4.31\pm0.74$ \\
461601491 & RGB & $1.77\pm0.02$ & $0.99$ & $1.9\pm0.2$ & $0.65\pm0.19$ \\
461620462 & RC & $4.01\pm0.13$ & $1$ & $1.38\pm0.18$ & $4.75\pm0.58$ \\
461620616 & RC & $3.90\pm0.08$ & $1$ & $1.45\pm0.13$ & $1.07\pm0.32$ \\
461627758 & RGB & $3.68\pm0.05$ & $0.97$ & $1.10\pm0.09$ & $2.02\pm0.83$ \\
\enddata
\end{deluxetable*}

\subsection{Mode Visibilities}
\label{sec:mode_visibilities}
Asteroseismology is a powerful tool to explore the internal physics of giant stars due to the presence of mixed modes, which is the result of coupling between the pressure and gravity modes. One branch of this research relates to the discovery a significant number of `dipole-suppressed' stars in samples of \textit{Kepler} red giants \citep{Mosser12_suppressedstars}. These stars exhibited reduced amplitudes in the $l=1$ mode relative to the radial modes. It has been proposed that the suppression mechanism is magnetic fields in the cores of these stars, which are induced by convection during the main sequence \citep{Fuller15_magneticfields}. This theory was supported with the evidence that the rate of dipole-suppressed stars increases with mass, and that there are no detections of suppressed stars below a mass of $1.1\msun$ \citep{Stello2016_suppressedNature}, which corresponds to the boundary between a radiative and convective core during the main sequence \citep{Kippenhahn_textbook}. To date, studying the rate of suppressed stars has mostly been limited to lower RGB stars in the \textit{Kepler} field. The exception is the Solar-metallicity open cluster M67, where \citet{Reyes25_M67} found a higher proportion of dipole-suppressed stars by a factor of $\sim6$ when compared to stars of a similar mass in the \textit{Kepler} field sample. They speculated that this could be related to the impact of cluster environments to the internal magnetic fields in stars, and emphasized that the mode suppression rate needs to be quantified in additional Galactic clusters. Thus, we extend this analysis by studying this phenomenon in NGC 188. 

We measure the mode visibilities of our NGC 188 gold sample following the method outlined in \citet{Stello16_PASAsuppressed}, where the oscillation spectrum is folded into four segments centred on $\numax$ of lengths equal to $\Delta\nu$. Using pre-defined frequency regions for the modes based on the estimate for radial mode phase offset ($\epsilon_0$), individual modes are identified in the original power spectrum. The dipole visibility ($V^2_{\ell=1}$) is defined as the ratio of the total integrated power of the peaks for a given non-radial mode to the total integrated power of radial modes.

To estimate the mode visibility uncertainties, we adopt a similar method as \citet{Crawford24}; the mode visibilities are recalculated for each individual \textit{TESS} sector and an uncertainty is taken as the standard error on the mean. This method captures the stochasticity of the mode amplitudes to give an estimate of the variation in the ratio between non-radial and radial modes vary over time. We note that the 27-day baseline of a single \textit{TESS} sector was too short to resolve each mode at these lower frequencies. Hence, the mode identification method often did not correspond to individual peaks in the spectrum, but rather blends of oscillation peaks. This led to a larger variation in calculated mode visibilities, and a larger scatter in the final distributions and associated uncertainties. 


Our measured dipole-mode visibilities are shown in the top panel of Figure~\ref{fig:ModeVis_v2}. We also provide example power spectra with the mode identifications in Appendix~\ref{sec:A3_ModeVisibilties}. Dipole-suppressed stars are classified to have a visibility that is less than the fiducial line (represented by the dashed line in Fig.~\ref{fig:ModeVis_v2}), which was empirically derived in \citet{Stello2016_suppressedNature}. This trend was calibrated for \textit{Kepler} field stars on the lower RGB (corresponding to $\numax>70~\mu$Hz). As such, we are extrapolating to lower $\numax$ values where the mode visibilities have not been well characterized. Due to the lack of a better diagnostic for mode suppression in our $\numax$ regime, we adopt this classification but note that the results should be interpreted with some caution. 

Given that both of our measured RGB and RC average masses for NGC 188 are close to the boundary between a radiative and convective core during the main sequence of $1.1~\msun$ \citep{Kippenhahn_textbook}, we expect that our stars should not display evidence of mode suppression. However, we identify one dipole-suppressed star, which corresponds to the most massive star in our sample (see purple star in Fig.~\ref{fig:NGC188_masses}). The \'echelle diagram and power spectrum for this star are provided in Appendix~\ref{sec:A3_ModeVisibilties}, where only a clear radial mode ridge can be detected. This potential evidence of dipole-mode suppression could suggest that this star is the product of a merger event during its main sequence lifetime. A merger of two low-mass stars would increase the mass of the core. This in turn increases the temperature and energy generation rate of the hydrogen burning, leading to a convective core and thus the environment to get mode suppression. We discuss this potential-merger candidate further in Section~\ref{sec:potential_merger}.


\begin{figure}[ht!]
\plotone{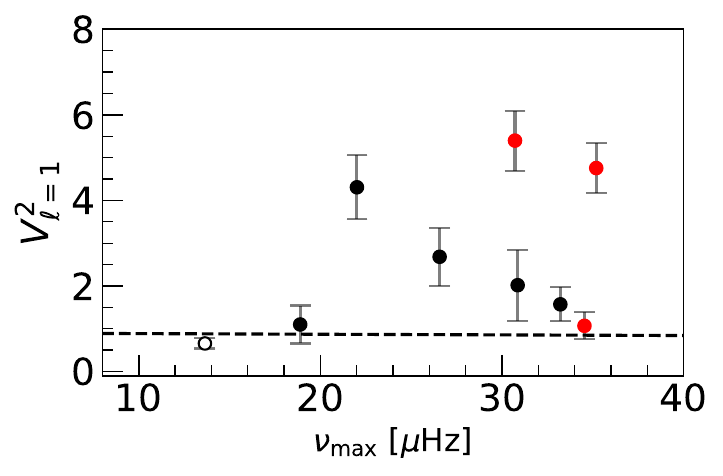}
\caption{Dipole-mode visibilities against $\numax$ for the NGC 188 gold sample. The dashed line in the top panel indicates the boundary between suppressed (below line) and normal stars (above line) from \citep{Stello2016_suppressedNature} extrapolated to lower $\numax$ values studied here. Points are color coded by evolutionary phase; RGB (black) and RC (red). The open circle indicates the potential dipole-suppressed star.
\label{fig:ModeVis_v2}}
\end{figure}

Unlike the analysis in \citet{Reyes25_M67} for M67, we do not observe an increase in the fraction of suppressed stars in NGC 188 compared to field stars of similar mass. However, our NGC 188 sample consists of luminous RGB and RC stars, whereas \citet{Reyes25_M67} limited their sample to RGB stars with $\numax > 70~\mu$Hz. Ultimately, to better quantify the measurement for the fraction of dipole-suppressed stars in NGC 188, we need to determine the mode visibilities for stars on the lower RGB (i.e. higher $\numax$ regime). At the moment, this is not possible with the current \textit{TESS} photometric data, and would require additional time-series observations. We also note that \citet{Stello16_PASAsuppressed} report an average dipole mode visibility of $V^2_{\ell=1} =1.5$, whereas we measure a median visibility of $V^2_{\ell=1} =2$. This larger value suggests that our measured dipole mode visibilities could be overestimated, likely due to the lower quality of the \textit{TESS} photometry compared to \textit{Kepler}. 


\subsubsection{TIC 461601491: Potential Merger Candidate}
\label{sec:potential_merger}
The RGB star TIC 461601491 has a seismic mass of $M_{\numax}=1.35\pm0.12~\msun$ using Eq.~\ref{eq:numax_mass_eq}, which is larger than the cluster average mass despite it not being classified as a statistical outlier (Fig.~\ref{fig:NGC188_masses}). Similar to our under-massive outliers, we rule out the possibility of this star being a non-member, as it passed both the astrometric and spectroscopic classifications. We speculate that the larger mass is a result of the star being a merger remnant for three reasons: i) potential evidence of mode-suppression, ii) the anomalous [C/N] abundance, and iii) variations in the time-series radial velocity measurements consistent with this star being in a binary system. We explore each of these further below.

As detailed in Sec.~\ref{sec:mode_visibilities}, it is thought that suppression of the dipole modes are due to magnetic fields induced by convection during the main sequence. This is unlikely to occur for NGC 188 stars with average masses close to the convective core boundary of 1.1$\msun$ \citep{Kippenhahn_textbook}. However, the merger of two low-mass stars during the main sequence that produced a more massive core could explain the observed mode suppression for TIC 461601491. If instead the star accreted mass onto its surface via a binary mass transfer event, the mass of the core would not increase. Therefore, the effect that causes mode suppression would not be expected. 

Furthermore, we can use the [C/N] abundance as a diagnostic of whether it gained mass prior to first dredge up or after this mixing event. As demonstrated in Figure~\ref{fig:CN_ages_masses}, TIC 461601491 has a [C/N] abundance that is significantly lower than the cluster average. From the [C/N]-mass trends in \citet{Roberts24_CN}, a lower [C/N] would equate to a larger stellar mass before the first dredge up, which is consistent with an binary interaction occuring prior to this mixing event. 

Finally, this star is classified as a single-lined spectroscopic binary with a period of $P=1241.7\pm1.1$ in the WOCS radial velocity survey \citep{Narayan25_WOCSRV}. As such, we propose that this star was originally in a triple system, where by the `Kozai-Lidov' mechanism \citep{Kozai1962,Lidov1962}, two of the companions collided during the main sequence forming the star TIC 461601491 and the third component remains in orbit around the merger remnant \citep[e.g.][]{Antognini16,Naoz16}. This star could be an interesting candidate for dynamical modeling to confirm if our proposed scenario is plausible.

\section{Summary \& Conclusions}
\label{sec:conclusion}

We measure the asteroseismology of red giants in two old metal-rich open clusters observed by the \textit{TESS} mission: NGC 6791 \& NGC 188. We use our own boutique analysis of the \textit{TESS} photometry, which involved careful individualized treatment of the aperture masks and light curves for each sector. We found that the SNR in our boutique power spectra are significantly larger compared to pipeline generated light curves. Following our results, we recommend that a boutique method of analyzing the photometry should be used for small \textit{TESS} samples, rather than adopting pipeline generated products.

For our NGC 6791 seismic sample, we extend previous \textit{TESS}-\textit{Kepler} comparisons by testing whether an increased observing baseline for \textit{TESS} photometry leads to improved accuracy in the measurement of $\numax$. Although similar oscillation mode frequencies are recovered from both datasets, we measure a larger median fractional offset in $\numax$ compared to previous studies. We attribute this reduced accuracy in the measurement of $\numax$ to the fainter magnitude of this cluster. As such, we do not report updated seismic parameters for our NGC 6791 sample. 

NGC 188 is significantly brighter, and thus has larger SNR in the \textit{TESS} power spectra. We present new seismic measurements for 17 red giants in NGC 188. Our main results from these NGC 188 measurements are summarised below:
\begin{itemize}
    \item We use our estimates for $\numax$, Gaia radii and spectroscopic $\Teff$ to derive seismic masses using Eq.~\ref{eq:numax_mass_eq}. From these masses, we estimate a median cluster mass of $M = 1.13\pm0.04$(rand)$^{+0.12}_{-0.19}$(sys)$~\msun$ for the RGB and $M=1.11\pm0.01$(rand)$^{+0.11}_{-0.19}$(sys)$~\msun$ for the RC, consistent with independent mass estimates from an isochrone and an eclipsing binary system. Furthermore, the precision on these average cluster masses are comparable to precisions on the seismic masses for \textit{Kepler} clusters and field stars. Our results highlight the value of a boutique method, and demonstrate that \textit{TESS} can yield excellent asteroseismic data with careful analysis.
    \item We estimate a difference between the RGB and RC median masses of $\Delta M = 0.02 \pm 0.04$(rand)$\pm0.01$(sys)$\msun$. Assuming an evolutionary initial mass differences of $0.01~\msun$, our measured mass loss would be $0.03~\msun$, which is smaller than the predicted mass loss from an isochrone of $0.07~\msun$ using a Reimers' efficiency of $\eta_{R} = 0.1$. Our value agrees with previous seismic mass loss measurements for open clusters at a similar metallicity. Our result contributes to the recent evidence of a dichotomy in the mass loss-metallicity trend between the metal-poor and metal-intermediate/rich populations, which cannot be described by the Reimers' mass loss scheme. 
    \item From our mass distributions, we discovered two cluster members (TIC 461593694 \& TIC 461626065) with masses that are statistically under-massive when compared to the cluster median mass for the RGB and RC. The RGB star has a [C/N] significantly larger than the cluster average, suggesting that it lost mass prior to the first dredge up event. We propose that these stars are the remnants of a post-common envelope mass stripping event. This scenario could be further investigated with binary population models.
    \item We use two methods to confirm the old age of NGC 188: i) fits to isochrones, and ii) [C/N]-age relation. By comparing our seismic masses and stellar parameters to isochrones, we infer seismic ages for our RGB sample. We measure a seismic median cluster age of $7.0\pm0.9~$Gyrs, which has a strong agreement with previous age determinations for this cluster. From the [C/N]-age relation, we derive a median cluster age of $6.3\pm0.5~$Gyrs. This again is consistent with previous literature values, and with our measured seismic age.
    \item We extend our asteroseismic analysis on a sub-sample of stars that were identified to display clear oscillation modes. For this `gold' sample of stars, we additionally measure $\Delta\nu$, and compare the seismic masses calculated using the $\numax$-only seismic relation (Eq.~\ref{eq:numax_mass_eq}) to the $\numax$-$\Delta\nu$ seismic relation (Eq.~\ref{eq:numax_dnu_mass_eq}). We find a discrepancy in the mass scales with a median fractional offset of 11\%. This offset is worse for our three RC stars, however this could be an effect of small-number statistics. We note that this disagreement between the mass scales could also be due to inaccuracies in measuring $\Delta\nu$. We emphasize that masses derived via the $\numax$-only seismic relation have been shown to be more robust for cluster seismic analyses, and as such are adopted as the final masses for our study.     
    \item By measuring the mode visibilities for our gold sample, we discovered one potentially dipole-suppressed star. Interestingly, this RGB star (TIC 61601491) has the largest mass of our NGC 188 sample ($M = 1.35\pm0.12~\msun$), has a [C/N] abundance that is lower than the cluster average, and has been identified as a spectroscopy binary. Given this evidence, we speculate that this star is the remnant of a merger between two main sequence stars. Dynamical modeling is needed to confirm our proposed scenario. 
\end{itemize}

\subsection{Future Work}
The all-sky \textit{TESS} mission has photometrically observed many more open clusters, where we have identified a further 10 clusters with at least 10 red giant asteroseismic candidates across the RGB and RC. By applying our boutique photometric pipeline to this \textit{TESS} data, we can replicate our analysis on clusters at different ages and metallicities. This will assist in investigations of the relationship between the RGB integrated mass loss and metallicities, and could resolve the behavior of the mass loss-metallicity trend between the metal-poor and metal-rich regimes. Specifically, for the metallicity range $-0.8\lesssim$~[M/H]~$\lesssim -0.5$, where the two trends have been shown to overlap but show a stark disagreement \citep{Li25_massloss}. There are two potential TESS open cluster candidates, NGC 2508 and NGC 2420, with [Fe/H]~$\sim-0.3$ that might provide a link between the two metallicity regimes, and help improve our poor understanding of the mass loss mechanism for low-mass stars. Finally, identifying binary remnants in the red giant phases of evolution across metallicities could contribute toward binary population studies, and help confirm the origins of blue stragglers and sub-sub giants progenitors in cluster environments.

 
While clusters are excellent stellar laboratories and are important benchmarks for asteroseismic analyses, they occupy discrete areas in metallicity-age space. To explore different Milky Way populations, it is necessary to study stars outside of clusters. The \textit{TESS} mission provides the unique opportunity to seismically study a broader and more diverse view of the Milky Way. The asteroseismic survey of $158,505$ red giants using one \textit{TESS} cycle \citep[][]{Hon21_tessrgbcatalog} has significantly increased the number of seismically-characterized stars compared to the previous APOKASC and APO-K2 catalogs. However, the full seven years of the \textit{TESS} mission has not yet been fully exploited; there exists the opportunity to further increase the sample of stars with new seismic measurements. Our study shows the promising results for accurately measuring the seismic masses and ages for \textit{TESS} red giants in clusters, which will be important benchmarks for future \textit{TESS} asteroseismic investigations. 

Furthermore, the up-coming \textit{Roman} infrared space mission \citep{Akeson19_Roman_telescope} will use high-resolution imaging to measure precise seismic parameters of bulge stars; a region of the Milky Way that has so far remained inaccessible to asteroseismic studies. It has been predicted that \textit{Roman}'s Galactic bulge time-domain survey will have an asteroseismic yield of $\sim300,000$ new red giants \citep{Weiss25}. Additionally, the Roman instrument's high spatial resolution and large field of view is ideally suited for the crowded environments of stellar clusters, and has the potential to extend the seismic analysis of clusters further. 

%


\begin{acknowledgments}
M.H., J.A.J, \& M.H.P. acknowledges support from the NASA grant 80NSSC24K0637. 
D.S. is supported by the Australian Research Council Discovery Project DP250104267.

The authors would like to thank Rafael Garc\'ia, Dinil Palakkatharappil, Savita Mathur \& Lina Borg at the Université Paris-Saclay \& Instituto de Astrofísica de Canarias for their immensely helpful discussions in the construction of the light curves and power spectra. We also appreciate the useful comments from those within the APOTESS collaboration. 

This paper includes data collected by the Kepler mission and obtained from the MAST data archive at the Space Telescope Science Institute (STScI). Funding for the Kepler mission is provided by the NASA Science Mission Directorate. This paper also includes data collected with the TESS mission, obtained from the MAST data archive at the Space Telescope Science Institute (STScI). Funding for the TESS mission is provided by the NASA Explorer Program. STScI is operated by the Association of Universities for Research in Astronomy, Inc., under NASA contract NAS 5–26555.

Some of the data presented in this paper were obtained from the Mikulski Archive for Space Telescopes (MAST) at the Space Telescope Science Institute. The specific observations analyzed can be accessed via \dataset[http://dx.doi.org/10.17909/t9-mrpw-gc07]{http://dx.doi.org/10.17909/t9-mrpw-gc07} \& \dataset[https://doi.org/10.17909/0cp4-2j79]{https://doi.org/10.17909/0cp4-2j79}. Support to MAST for these data is provided by the NASA Office of Space Science via grant NAG5–7584 and by other grants and contracts.

This research made use of Lightkurve, a Python package for Kepler and TESS data analysis \citep{LightKurve2018}.

Funding for the Sloan Digital Sky Survey V has been provided by the Alfred P. Sloan Foundation, the Heising-Simons Foundation, the National Science Foundation, and the Participating Institutions. SDSS acknowledges support and resources from the Center for High-Performance Computing at the University of Utah. SDSS telescopes are located at Apache Point Observatory, funded by the Astrophysical Research Consortium and operated by New Mexico State University, and at Las Campanas Observatory, operated by the Carnegie Institution for Science. The SDSS web site is \url{www.sdss.org}.

SDSS is managed by the Astrophysical Research Consortium for the Participating Institutions of the SDSS Collaboration, including the Carnegie Institution for Science, Chilean National Time Allocation Committee (CNTAC) ratified researchers, Caltech, the Gotham Participation Group, Harvard University, Heidelberg University, The Flatiron Institute, The Johns Hopkins University, L'Ecole polytechnique f\'{e}d\'{e}rale de Lausanne (EPFL), Leibniz-Institut f\"{u}r Astrophysik Potsdam (AIP), Max-Planck-Institut f\"{u}r Astronomie (MPIA Heidelberg), Max-Planck-Institut f\"{u}r Extraterrestrische Physik (MPE), Nanjing University, National Astronomical Observatories of China (NAOC), New Mexico State University, The Ohio State University, Pennsylvania State University, Smithsonian Astrophysical Observatory, Space Telescope Science Institute (STScI), the Stellar Astrophysics Participation Group, Universidad Nacional Aut\'{o}noma de M\'{e}xico, University of Arizona, University of Colorado Boulder, University of Illinois at Urbana-Champaign, University of Toronto, University of Utah, University of Virginia, Yale University, and Yunnan University. 

\end{acknowledgments}

\begin{contribution}

M.H. led the analysis and writing of the manuscript. J.A.J. and M.H.P. oversaw the project's progress and provided regular feedback on the analysis. L.M. derived seismic ages for the sample. J.T., J.D.R., D.S. and M.M. provided feedback on the manuscript and occasional feedback on the analysis.

\end{contribution}

%
\facilities{TESS, Kepler, Gaia, Sloan (APOGEE), WIYN}

\software{NumPy \citep{Harris20_Numpy}, astropy \citep{2013A&A...558A..33A,2018AJ....156..123A,2022ApJ...935..167A}, SciPy \citep{Virtanen20_scipy}, Matplotlib \citep{Hunter_Matplotlib}, LightKurve \citep{LightKurve2018}, TESSCut \citep{Brasseur2019_tesscut}, pyMON \citep{Howell25_pyMON}, pySYD \citep{Chontos21_pySYD}, tess-point \citep{Burke20_tesspoint}, Asfgrid \citep{Sharma2016_asfgridpaper,Stello2022_asfgrid_v2}, Pandas \citep{Pandas25}, isochrones \citep{Morton15_isochronepython}, kiauhoku \citep{Claytor2020_kiahoku}, echelle \citep{daniel_hey_2020_3629933}, ChatGPT \citep{openai2025chatgpt}.}


\appendix

\section{Tables of Results for NGC 188 \& NGC 6791}
\label{sec:A1_ParamTables}

We provide truncated tables for stellar and seismic parameters used in this study for NGC 188 (Table~\ref{tab:NGC188_table}) and NGC 6791 (Table~\ref{tab:NGC6791_table}). Full tables with all stellar and seismic parameters can be accessed at this Zenodo repository: 10.5281/zenodo.19337401.

\begin{deluxetable*}{lccccccccc}
\digitalasset
\tablewidth{0pt}
\tablecaption{Results for our NGC 188 red giant sample. This includes the identification in the \textit{TESS} input catalog (TIC), evolutionary phase classification (Evol), identification in the WIYN open cluster study (WOCS; \citealt{Narayan25_WOCSRV}), the number \textit{TESS} sectors ($N_{\text{TESS}}$) included in our light curves, and the measured $\numax$, $\numax$ correction ($f_{\numax}$), $\Teff$, Gaia radius, seismic mass from Eq.~\ref{eq:numax_mass_eq}, and seismic age. The $^*$ indicates that spectroscopic information was taken from the Gaia XP XGBOOST catalog \citep{Andrae2023_GaiaXGBoost}, rather than APOGEE DR17 (see Sec.~\ref{sec:target_selection}).  Additional parameters will be made available in an online catalog. \label{tab:NGC188_table}}
\tablehead{
\colhead{TIC ID} &	\colhead{Evol}	 	& \colhead{WOCS ID}	& \colhead{$N_{\text{TESS}}$} &	\colhead{$\numax$ ($\mu$Hz)}	& \colhead{$f_{\numax}$} & \colhead{$R_{\text{Gaia}}$ ($\rsun$)} & \colhead{$\Teff$ (K)} &	\colhead{$M_{\numax}$ ($\msun$)}	 & \colhead{Seismic Age (Gyrs)}}
\startdata
461595568                                             & RGB                             & 3015                                                   & 10                                           & 33.3$\pm$0.4                                  & 1.00                                                      & 10.5$\pm$0.4                                           & 4507                                   & 1.05$\pm$0.08                             & 10.1$\pm$2.6                                  \\
461597443                                             & RC                              & 3271                                                   & 10                                           & 34.7$\pm$0.4                                  & 0.99                                                      & 10.5$\pm$0.4                                           & 4673                                   & 1.12$\pm$0.08                             &                                               \\
461590328*                                             & RGB                             & \multicolumn{1}{l}{}                                   & 11                                           & 59.7$\pm$0.6                                  & 1.00                                                      & 7.5$\pm$0.3                                            & 4619                                   & 0.97$\pm$0.07                             & 13.5$\pm$4.0                                  \\
461593694                                             & RGB                             & \multicolumn{1}{l}{}                                   & 10                                           & 26.6$\pm$0.4                                  & 1.01                                                      & 9.7$\pm$0.3                                            & 4612                                   & 0.73$\pm$0.06                             &                                               \\
461601193                                             & RC                              & 5085                                                   & 10                                           & 33.0$\pm$0.5                                  & 0.99                                                      & 10.8$\pm$0.4                                           & 4650                                   & 1.10$\pm$0.08                             &                                               \\
461599427                                             & RC                              & 4150                                                   & 11                                           & 30.9$\pm$0.4                                  & 0.99                                                      & 11.1$\pm$0.4                                           & 4661                                   & 1.11$\pm$0.08                             &                                               \\
461601138*                                             & RGB                             & 5887                                                   & 10                                           & 18.4$\pm$0.4                                  & 1.02                                                      & 15.3$\pm$0.6                                           & 4386                                   & 1.23$\pm$0.10                             & 5.7$\pm$1.5                                   \\
461601229                                             & RGB                             & 4668                                                   & 10                                           & 22.0$\pm$0.2                                  & 1.01                                                      & 13.4$\pm$0.5                                           & 4437                                   & 1.13$\pm$0.09                             & 7.0$\pm$1.9                                   \\
461601491                                             & RGB                             & 4843                                                   & 11                                           & 13.4$\pm$0.4                                  & 1.03                                                      & 18.7$\pm$0.8                                           & 4387                                   & 1.35$\pm$0.12                             & 4.1$\pm$1.3                                   \\
461618921                                             & RGB                             & 5835                                                   & 11                                           & 38.2$\pm$0.8                                  & 1.00                                                      & 9.9$\pm$0.4                                            & 4534                                   & 1.07$\pm$0.08                             & 9.3$\pm$2.3                                   \\
461620462                                             & RC                              & 9159                                                   & 10                                           & 35.2$\pm$0.5                                  & 0.99                                                      & 10.7$\pm$0.4                                           & 4680                                   & 1.17$\pm$0.09                             &                                               \\
461620563                                             & RC                              & 6353                                                   & 10                                           & 32.7$\pm$0.9                                  & 0.99                                                      & 10.9$\pm$0.4                                           & 4631                                   & 1.12$\pm$0.09                             &                                               \\
461620616                                             & RC                              & 6982                                                   & 10                                           & 34.4$\pm$0.5                                  & 0.99                                                      & 10.5$\pm$0.4                                           & 4673                                   & 1.09$\pm$0.08                             &                                               \\
461620688                                             & RGB                             & 5597                                                   & 11                                           & 51.9$\pm$1.0                                  & 1.00                                                      & 9.1$\pm$0.3                                            & 4588                                   & 1.23$\pm$0.09                             & 5.9$\pm$1.6                                   \\
461625776                                             & RGB                             & 6687                                                   & 12                                           & 5.7$\pm$0.1                                   & 1.06                                                      & 25.8$\pm$1.2                                           & 4100                                   & 1.09$\pm$0.11                             & 7.9$\pm$2.7                                   \\
461626065                                             & RC                              & 1947                                                   & 11                                           & 32.5$\pm$0.3                                  & 0.99                                                      & 9.7$\pm$0.4                                            & 4662                                   & 0.88$\pm$0.07                             &                                               \\
461627758                                             & RGB                             & 6602                                                   & 11                                           & 30.9$\pm$0.6                                  & 1.00                                                      & 11.4$\pm$0.4                                           & 4512                                   & 1.15$\pm$0.09                             & 6.5$\pm$2.1                                   \\         
\enddata
\end{deluxetable*}

\begin{deluxetable*}{lcccc}
\digitalasset
\tablewidth{0pt}
\tablecaption{Results for our NGC 6791 red giant sample. We provide the identification in the \textit{Kepler} input catalog (KIC). See Table~\ref{tab:NGC188_table} for further description of columns. \label{tab:NGC6791_table}}
\tablehead{
\colhead{TIC ID}& \colhead{KIC ID} & \colhead{Evol} & \colhead{$N_{\text{TESS}}$} & \colhead{$\numax$ ($\mu$Hz)}}
\startdata
122374305                                             & 2298436                                               & RC                              & 4                                                                & 27.48$\pm$0.63                      \\
122364738                                             & 2437804                                               & RC                              & 2                                                                & 22.35$\pm$1.13                      \\
122364375                                             & 2437965                                               & RGB                             & 7                                                                & 8.45$\pm$0.17                       \\
122364981                                             & 2437164                                               & RC                              & 3                                                                & 27.18$\pm$0.31                      \\
122364934                                             & 2437340                                               & RGB                             & 6                                                                & 8.35$\pm$0.26                       \\
122364528                                             & 2437805                                               & RC                              & 2                                                                & 27.37$\pm$0.67                      \\
122364384                                             & 2437496                                               & RGB                             & 6                                                                & 4.70$\pm$0.09                       \\
122364589                                             & 2436884                                               & RGB                             & 6                                                                & 8.81$\pm$0.40                    \\
122364162                                             & 2569935                                               & RGB                             & 7                                                                & 5.72$\pm$0.28                      \\
122364229                                             & 2569360                                               & RGB                             & 5                                                                & 22.97$\pm$0.42                      \\
122364081                                             & 2569204                                               & RGB                             & 6                                                                & 5.30$\pm$0.24                       \\
122375300                                             & 2708270                                               & RGB                             & 2                                                                & 15.69$\pm$0.41                      \\
\enddata
\end{deluxetable*}

\section{Extended Discussion on $f_{\nu_{\text{max}}}$ Corrections}
\label{sec:A2_fnumax_corr}

The APOKASC-3 catalog estimated the $f_{\numax}$ corrections by invoking that the seismic radius (see their Eq. 2) must equal the Gaia radius. This was completed for three different models that calculate $f_{\Delta\nu}$, which is an input in the seismic radius. For each model, they empirically derive the zero-point $f_{\numax}$ corrections for the RC and RGB stars with $\nu_{\text{max}} > 50~\mu$Hz separately. For the luminous RGB stars, a polynomial as a function of $\nu_{\text{max}}$ was provided as a non-linear ratio was found between the seismic and Gaia radii (see their Fig.~8). Again, we emphasize that APOKASC-3 define $f_{\numax}$-corrections as the reciprocal of Eq.~\ref{eq:numax_mass_eq}, hence these corrections are always implemented as $f_{\numax}^{-1}$ for our investigation. 

Table~\ref{tab:average_masses_different_models} shows a summary of the average masses when using the $f_{\numax}$ corrections from the three different models. We also test the effect of assuming $f_{\numax}=1$. We have not removed any outliers from these average estimates. We see consistency between the average masses for each method, and conclude that the choice of the $f_{\numax}$ correction scheme does not have any significant systematic effects to the average cluster masses. We adopt the GARSTEC+Mosser model for our NGC 188 seismic masses, as it was the same method that was used for the APOKASC-3 catalog.

\begin{deluxetable*}{lcc}
\digitalasset
\tablewidth{0pt}
\tablecaption{Comparison of the average masses calculated using different $f_{\numax}$ corrections schemes from the APOKASC-3 catalog. Mass outliers have not been removed for these average measurements. \label{tab:average_masses_different_models}}
\tablehead{
\colhead{$f_{\numax}$ Correction Schemes} & \colhead{$M_{\text{RGB}}$ ($M_{\odot}$)} &  \colhead{$M_{\text{RC}}$ ($M_{\odot}$)} }
\startdata
$f_{\numax}=1$ & $1.13\pm0.03$ & $1.13\pm0.04$ \\
Sharma+White  & $1.13\pm0.04$ & $1.12\pm0.04$ \\
GARSTEC+Mosser  & $1.14\pm0.04$ & $1.12\pm0.04$  \\
GARSTEC+White  & $1.15\pm0.04$ & $1.13\pm0.04$\\
\enddata
\end{deluxetable*}

\section{Extended Discussion on Mode Visibilities}
\label{sec:A3_ModeVisibilties}

\begin{figure*}[ht!]
\plotone{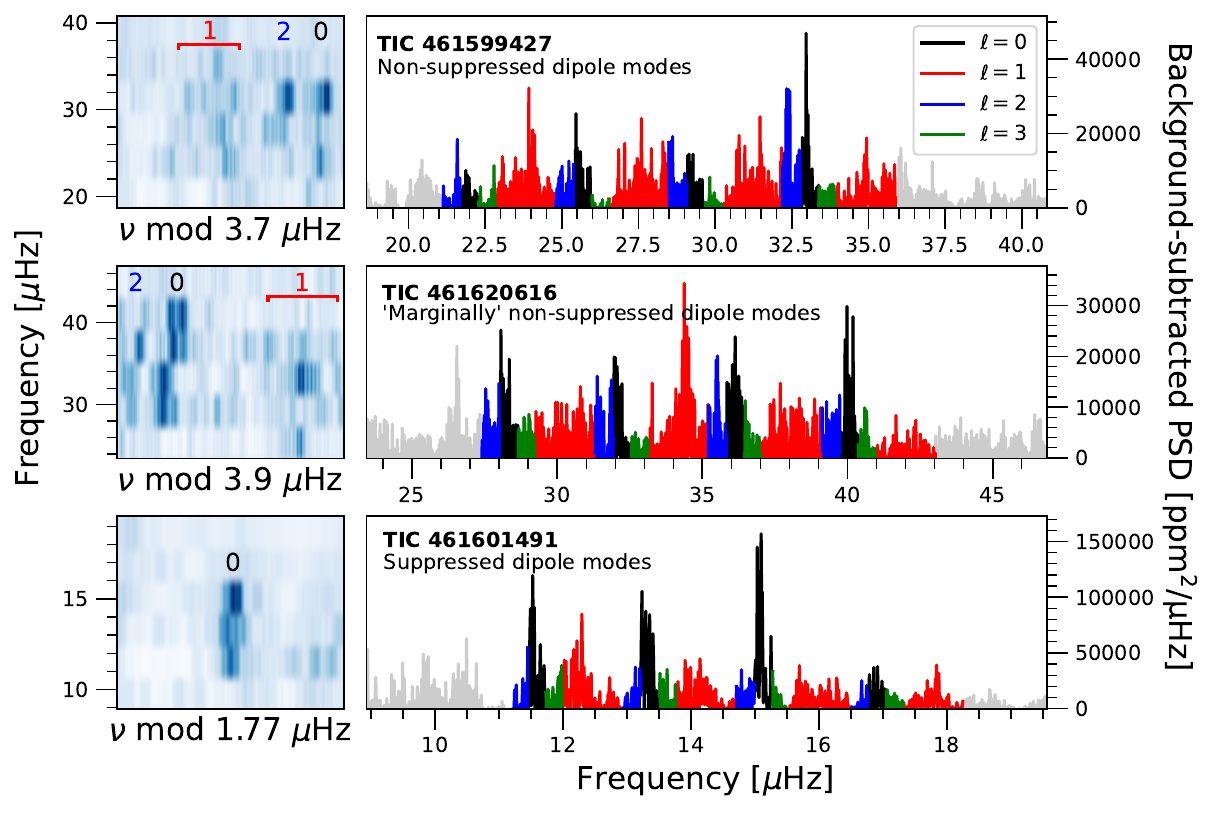}
\caption{\'Echelle diagrams (left) and power spectra (right) for a non-suppressed dipole mode star (top), `marginally' non-suppressed dipole mode star (middle), and suppressed dipole mode star (bottom). The ridges in the \'echelle diagram are labeled by the spherical harmonic $\ell$ degree. The colors in the power spectra illustrate the mode identification for four radial orders of the power excess. 
\label{fig:PSD_MV}}
\end{figure*}

In Section~\ref{sec:mode_visibilities}, we discuss our methodology to estimate the dipole-mode visibilities for our gold sample in NGC 188. Here we provide three example power spectra and the associated \'echelle diagrams (using the \texttt{echelle} software; \citealt{daniel_hey_2020_3629933}) that illustrate the mode identification for a non-suppressed, marginally non-suppressed (could be defined as suppressed from 1$\sigma$ uncertainties), and suppressed dipole stars in Figure~\ref{fig:PSD_MV}. An échelle diagram is constructed by stacking segments of the power spectrum of length $\Delta\nu$, where modes of the same degree $\ell$ align into distinct ridges. In the \'echelle diagrams for the marginally non-suppressed and non-suppressed stars, both the radial ridge (labeled with a `0') and ridges corresponding to non-radial modes ($\ell>0$; also denoted) are clearly observed. This indicates that are `marginally' non-suppressed star is in fact a true non-suppressed star, despite exhibiting a lower dipole-mode visibility. Whereas in the \'echelle diagram for the non-suppressed star, we only detect a clear radial mode ridge, supporting its classification as a likely dipole-suppressed star.


\bibliography{refs}{}
\bibliographystyle{aasjournalv7}



\end{document}